\documentclass[12pt]{article}

\usepackage{amssymb,amsmath,amsfonts,eurosym,geometry,ulem,color,setspace,sectsty,comment,footmisc,pdflscape,array,hyperref}

\usepackage[
    backend=biber,
    style=numeric,
    sorting=none,
    giveninits=true,
    maxbibnames=3,
    minbibnames=3,
    maxcitenames=2,
    date=year,
    url=false,
    doi=false
]{biblatex}
\DeclareBibliographyDriver{book}{%
  \usebibmacro{bibindex}%
  \usebibmacro{begentry}%
  \usebibmacro{author/editor+others/translator+others}%
  \setunit{\labelnamepunct}\newblock
  \usebibmacro{maintitle+title}%
  \newunit
  \printlist{language}%
  \newunit\newblock
  \usebibmacro{byauthor}%
  \newunit\newblock
  \usebibmacro{byeditor+others}%
  \newunit\newblock
  \printfield{edition}%
  \newunit
  \iffieldundef{maintitle}
    {\printfield{volume}%
     \printfield{part}}
    {}%
  \newunit
  \printfield{volumes}%
  \newunit\newblock
  \usebibmacro{series+number}%
  \newunit\newblock
  \printfield{note}%
  \newunit\newblock
  \printlist{organization}%
  \newunit
  \usebibmacro{publisher+location+date}
  \newunit\newblock
  \usebibmacro{chapter+pages}%
  \newunit
  \printfield{pagetotal}%
  \newunit\newblock
  \iftoggle{bbx:isbn}
    {\printfield{isbn}}
    {}%
  \newunit\newblock
  \usebibmacro{doi+eprint+url}%
  \newunit\newblock
  \usebibmacro{addendum+pubstate}%
  \setunit{\bibpagerefpunct}\newblock
  \usebibmacro{pageref}%
  \newunit\newblock
  \usebibmacro{related}%
  \usebibmacro{finentry}}

\DeclareFieldFormat[book,mvbook,collection,mvcollection,booklet]{date}{[#1]}
\DeclareFieldFormat[book,mvbook,collection,mvcollection,booklet]{origdate}{[#1]}
\DeclareFieldFormat[book,mvbook,collection,mvcollection,booklet]{eventdate}{[#1]}
\DeclareFieldFormat[book,mvbook,collection,mvcollection,booklet]{urldate}{[#1]}

\usepackage{subcaption}
\usepackage{graphicx}
\usepackage{amsmath}
\usepackage{amsthm}
\usepackage{amssymb}
\usepackage{multirow}
\usepackage{url}
\usepackage{authblk}

\newtheorem{Definition}{Definition}

\newtheorem*{Proof}{Proof}
\newtheorem{proposition}{Proposition}

\newcolumntype{L}[1]{>{\raggedright\let\newline\\arraybackslash\hspace{0pt}}m{#1}}
\newcolumntype{C}[1]{>{\centering\let\newline\\arraybackslash\hspace{0pt}}m{#1}}
\newcolumntype{R}[1]{>{\raggedleft\let\newline\\arraybackslash\hspace{0pt}}m{#1}}

\begin{document}

\begin{titlepage}
\title{Evolutionary Cooperation with Game Transitions via Markov Decision Chain in Networked Population\thanks{This work was supported in part by the Natural Science Foundation of Chongqing under Grant CSTB2025YITP-QCRCX0007; in part by the National Nature Science Foundation of China (NSFC) under Grant 62206230; and in part by the National Research, Development and Innovation Office (NKFIH) under Grant K142948.}}

\author[1]{Chaoyang~Luo}
\author[1]{Yuji~Zhang}
\author[1]{Minyu~Feng\thanks{Corresponding author. Email: \texttt{myfeng@swu.edu.cn}}}
\author[2]{Attila~Szolnoki}

\affil[1]{College of Artificial Intelligence, Southwest University, Chongqing 400715, China}
\affil[2]{Institute of Technical Physics and Materials Science, Centre for Energy Research, P.O. Box 49, Budapest H-1525, Hungary}

\date{}

\maketitle
\begin{abstract}
    \noindent Individual cooperative strategy influences the surrounding dynamic population, which in turn affects cooperative strategy. To better model this phenomenon, we develop a Markov decision chain based game transitions model and examine the dynamic transitions in game states of individuals within a network and their impact on the strategy's evolution. Additionally, we extend single-round strategy imitation to multiple rounds to better capture players' potential non-rational behavior. Using intensive simulations, we explore the effects of transition probabilities and game parameters on game transitions and cooperation. Our study finds that strategy-driven game transitions promote cooperation, and increasing the transition rates of Markov decision chains can significantly accelerate this process. By designing different Markov decision chains, these results provide simulation based guidance for practical applications in swarm intelligence, such as strategic collaboration.
\vspace{0in}
\bigskip
\end{abstract}
\setcounter{page}{0}
\thispagestyle{empty}
\end{titlepage}
\pagebreak \newpage

\doublespacing

\section{Introduction}

    Evolutionary game theory is a fundamental framework for studying the mechanisms behind cooperative behavior among organisms in social dilemmas. Social dilemmas arise when there is a conflict between maximizing personal benefits and promoting group benefits. In such a situation, participants who cooperate often take a risk to enhance the overall benefits of the group~\cite{Ohtsuki2006}. However, cooperation is widespread, from mutualistic interactions among animals in nature to collaborative behaviors in human societies~\cite{zhang2024limitation}. During the past two decades, the study of dynamic processes in complex networks~\cite{zeng2025bursty, zeng2025complex}, such as epidemics~\cite{tanimoto2021sociophysics}, community development~\cite{Pi2025Dynamic}, and the motivations and mechanisms for individuals to cooperate in social dilemmas, have gained increasing importance and became a focal point of contemporary research~\cite{Li2026, Deonauth2021, Xiao2023}.

    In games, the state of the environment may change due to the consequence how individuals' vary their strategies~\cite{Hilbe2018, Su2019, kleshnina2023effect}. This observation was already reported in studies how bacterial population fitness changes~\cite{tilman2020evolutionary}, the balance of animal social behaviors, and the sustainable management of public resources~\cite{menge2008evolutionary, grman2012ecological}. In previous research, by describing the direct impact of strategies on game payoffs, different strategy combinations of individuals in a population can directly alter the game state, known as stochastic games or game transitions~\cite{Hilbe2018,Su2019}. Interestingly, using Markov models to depict the dynamic process of environmental changes has been proven a natural approach~\cite{Feng2023}. However, it is important to note that despite significant progress in existing studies, the majority of proposed models assume that individuals only play in a single-game or fixed-game system, and can quickly update their strategies via an imitation at each time step~\cite{su2025evolutionary, tarnita2025reconciling}. To address this limitation, scholars have proposed various improvement directions. For instance, introducing multiplayer games with nonlinear payoff functions to enhance model complexity~\cite{Perc2012, Aming2016} and considering the evolution of cooperation in stochastic games as an extension of static game states~\cite{Hilbe2018}. Alternatively, the construction of evolutionary dynamics models with game transitions can simulate the dynamic interaction between individual behaviors and environmental conditions. Studies have shown that such game transition mechanisms can effectively reduce the critical benefit-to-cost ratio of cooperation, thereby promoting its emergence~\cite{Su2019}. In addition, memory-based Snowdrift Game (SDG) models incorporate bounded rationality assumptions~\cite{wang2006}, where individuals' decisions rely on memories of past strategies, enabling increased cooperation levels even when defectors occupy hubs in a heterogeneous graph. As an important step, Feng {\it et al.} proposed a game transition model based on Markov chains~\cite{Feng2023}, which better aligns with environmental randomness. Their model assumes that game state transitions follow a purely random process governed by a single Markov chain, failing to consider the dependence of state transitions on individuals' strategies. To this end, this paper will introduce Markov decision chains, allowing different strategies to drive corresponding Markov chains, thereby guiding the environment to develop in different directions while modeling the degree of individuals' irrationality. In our model,  strategies affect the transition probabilities rather than imposing immediate switches. We analyze the stationary cooperation level under different environmental configurations.
	
    When individuals’ decisions affect the game environment, it is evident that strategies that benefit the environment tend to prevail. However, this conclusion does not always hold, particularly when environmental changes are subtle or when the environment is difficult to improve through individual decisions. Therefore, it is necessary to model this process and investigate its effects numerically, especially regarding the rate of environmental change, which represents the extent to which strategies influence the environment and serves as a threshold indicator for the impact on both the environment and the spread of strategies. We stress that past models have not been sufficiently specific in modeling how game transitions interact with individual strategies. Some models chose to ignore this and treated game transitions as independent Markov chains~\cite{Feng2023}, while others established strategy dependencies that did not adequately model the transition process of the environment~\cite{Hilbe2018}. Although simplification can yield favorable mathematical properties, it weakens the description of environmental changes. Therefore, we now model individual decision-making processes, such as whether to cooperate or defect, as a Markov decision chain. This implies that as strategies evolve, the game states will no longer follow a single independent Markov chain. Instead, different strategies will induce different dynamics of game transitions. Individuals select strategies at each state, impacting both their payoffs and transition probabilities between games, which in turn shapes the future game state. As a result, when the Markov chains driven by cooperation and defection are identical, our model reduces to a simple Markov chain-based game transition model. When cooperation and defection no longer govern the Markov chain but directly alter the game state, it reduces to a stochastic game model. Our model represents a potential generalization of previous models. Then we expand on the traditional single-round strategy update rule and allow individuals to update multiple strategies within a single time step, reflecting the reality that strategy decisions are often made over an extended period.

    Summing up, our work has produced the following innovative results and potential applications:
    \begin{itemize}
        \item It is the first attempt to introduce a Markov decision chain to model the strategy-driven game transition process. Building upon previous uses of Markov chains to describe game transitions, this approach incorporates the dependence of transitions on strategies, while also characterizing the degree of individual irrationality.
        \item Through extensive simulations, we identify new insights into how environmental feedback influences the emergence of cooperation. In particular, when environmental transitions are not directly determined by strategies, strategies can instead regulate the environment indirectly by affecting the transition rates. Our results show that increases in both the rate and the magnitude of environmental changes significantly promote higher levels of cooperation.
        \item Our findings provide a new model and simulation-based insights into the interactions between strategies and environments, particularly in situations where environmental changes are regulated by specific transition rates rather than being directly dictated by strategies. Results offer guidance for formulating public cooperation policies, and serve as a reference for coordination optimization in multi-agent systems.
    \end{itemize}
    The structure of this paper is as follows: we first introduce our model in terms of game transitions and evolutionary dynamics in Sec.~\ref{Model}. Subsequently, Sec.~\ref{Simu} presents the results of our simulations to explore a range of properties of the model. Finally we provide a summary and conclusion of our research in Sec.~\ref{Conclu}.

	\section{Evolutionary Game Modelling}		\label{Model}
	
    Previous research on game transitions over general Markov chains does not accurately capture the unique impact of cooperation behavior on these transitions in networked populations~\cite{Hilbe2018, Feng2023}.    
    The Markov property in our model has been supported by previous studies~\cite{Vasconcelos2017}. In populations, an individual's prior decisions can affect subsequent payoffs or the dynamics of future game iterations~\cite{Franzenburg2013, Simon2014, Acar2008}. As an example, in a social exchange, such as mutual gifting, lack of a reciprocal gift can reduce the perceived value of future gifts. Consistent with previous game transition models~\cite{Su2019}, we assume that the strategies of agents influence and are influenced by their current game states. The key distinction is that previous studies typically assumed game transitions to be directly determined by strategies. In contrast, we introduce a Markov chain, where strategies modify the transition rates and thereby regulate game transitions indirectly.
           
    We therefore propose a Markov decision process based on a bidirectional feedback between game state transitions and strategies. Specifically, agents in the network do not apply strategies in a fixed game state but in dynamic game states that transition according to their strategies. Meanwhile, changes in game states further influence the strategies of agents, forming a self-regulating feedback loop. This model can be used to describe scenarios when the collective strategy choice affects the environment, which in turn influences agents on how to adjust their strategies.
	
	\subsection{Transition Rules for Game States}
	
    In this subsection, we introduce the transition rules for game states. First, we assume that game transitions follow a Markov decision process with a finite state space $G=\{ G_1, G_2, \dots, G_n \}$, where $G_i$ represents different game states. These states correspond to different parameterized versions of a game model, such as PDG, SDG, and Donation Game, among others, and can also correspond to combinations of these games.

    The transition probabilities of the Markov decision chain are represented by the set $\{P, P^\prime\}$, implying that cooperative and defective behaviors can lead to different state transitions of a game. To give real-life examples, actions such as overfishing and overgrazing, which constitute defection, can render the entire environment system unfavorable to cooperation~\cite{Rankin2007}. In the opposite case, cooperation fosters a more conducive environment system. In our model, each agent performs a simulated game before engaging in the actual game. During the simulated game, the players compare their payoffs with their neighbors to plan their strategy for the next round. The selected strategy is reserved for a policy that guides future decision-making. The simulated games are maintained until the policy is completed. Subsequently, the players sequentially follow their reserved strategies in the actual games. As the game progresses, agents adjust the transition probabilities dynamically based on their chosen strategies, creating a feedback loop that influences future strategies and game states. This mechanism leads to complex behavioral patterns in evolutionary games, and the policy is defined as $U=\{\xi_1, \xi_2, \dots, \xi_n\}$, where $\xi_i \in \Theta = \{ C, D \}$, where $C$ represents cooperation and $D$ defection. 
    
    We further assume that the transition probabilities of an agent's game state are influenced by their current strategies. Specifically, if the agent chooses cooperation $(C)$ at the current moment, then their game states in the next moment will transition according to the probability matrix $P$; if the agent chooses defection $(D)$, their game states will transition according to the probability matrix $P^\prime$. For instance, if an agent starts from game state $G_1$ and the policy of this agent is $\{C, D, C\}$, then the agent's game states will transition according to the matrices $\{P, P^\prime, P\}$ in the three subsequent time steps. After that, the agent will update its policy.

	\subsection{Evolutionary Rule and Policy Making}
	
\begin{figure}
		\centering
		\includegraphics[width=\textwidth]{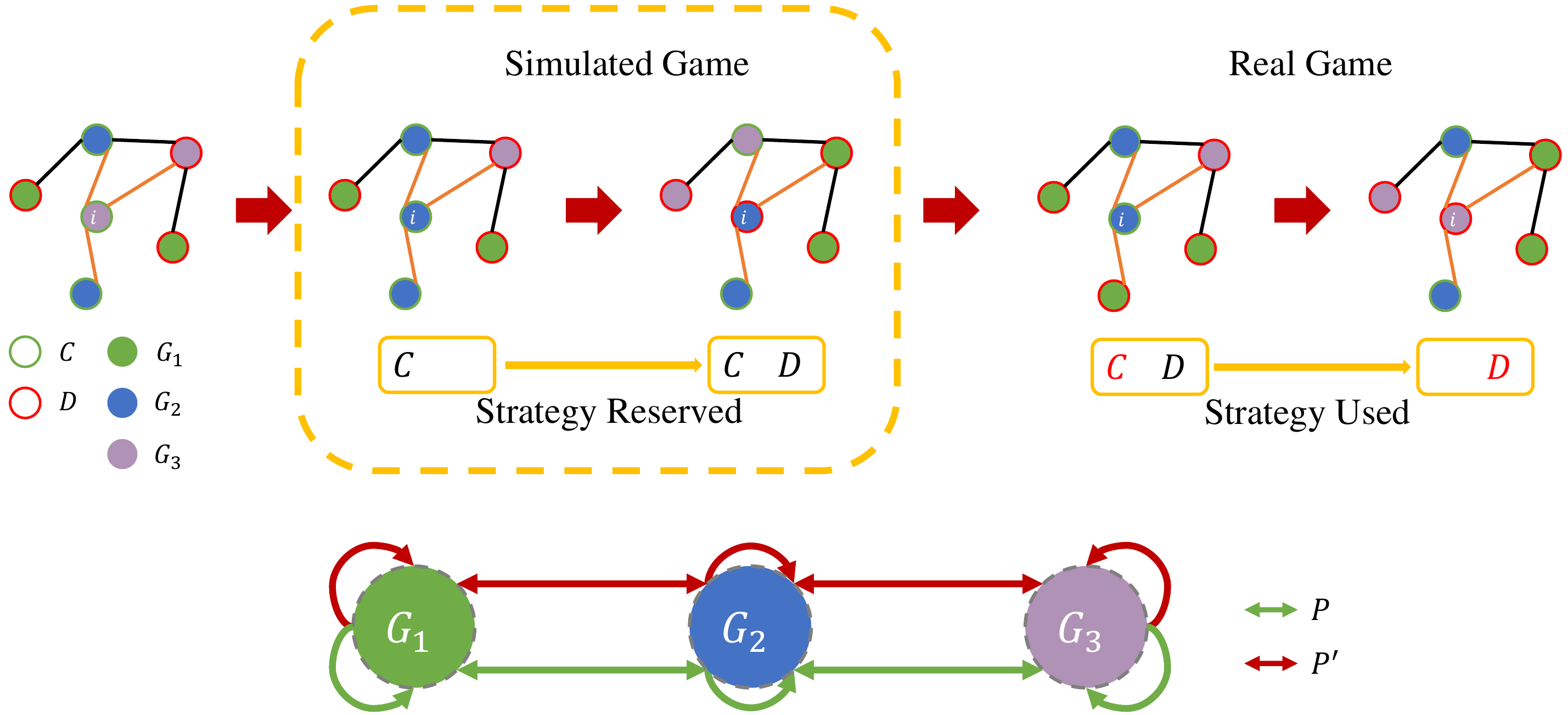}
            \caption{\textbf{Markov decision evolutionary model with simulated game.} This figure illustrates a possible example of strategy and state evolution, involving a Markov process composed of three game states $G_1$, $ G_2$, and $ G_3$. The green and red arrows represent the transition probabilities under cooperation and defection strategies, respectively. Taking agent $i$ as an example, this agent first engages in a simulated game to reserve strategies until its policy is fully constructed. Subsequently, in the real game, the agent will apply strategies based on the policy.}
		\label{fig:fig1}
	\end{figure}

    In networked populations, agents are not limited to apply a singular strategy at each moment; instead, they may plan multiple future strategies at a given time step. For instance, in the context of natural ecosystems, animals engaging in food storage behavior often devise strategies that account for future foraging needs. In our model, when agents update their policies, they first engage in a series of simulated games with their neighbors. In each round of simulated games, agents get a strategy based on the current game situation and reserve it in their policies. The simulated games continue until the policies are fully populated, after which these strategies in policies will be used in the subsequent real games. We illustrate an example in Fig.~\ref{fig:fig1} to visualize the details of the model, including the transition of game states and the evolution of strategies in a networked population. Notably, the strategies of node $i$ are not updated continuously, but periodically with multiple updates at a certain point. That is, at specific time points, multiple strategies are concurrently updated and reserved within a policy, to be employed in subsequent interactions. It is important to stress that an agent’s strategy influences the transition probabilities of its game states, while different game states in turn affect the agent's strategy updates, thus forming a self-referential bidirectional feedback loop.

    During each round of the simulated game, agents accumulate payoffs by playing games with all their neighbors. Specifically, if agent $j$ is in game state $G_i$, then mutual cooperation (defection) with a neighbor results in payoff $R_i$ ($P_i$) for both. In the case of unilateral cooperation, the cooperator receives $S_i$ while the defector receives $T_i$. Based on this, the payoffs $\Pi_{j}$ for agent $j$ can be represented by the following payoff matrix:
\begin{equation}
\Pi_{j}=\left(\begin{array}{ll}
	R_{i} & S_{i} \\
	T_{i} & P_{i}
\end{array}\right)\,.
\end{equation}

    After each round of the simulated game, agent $j$ compares the accumulated total payoff with all their neighbors and selects a neighbor $k$ whose accumulated payoff is higher than the accumulated total payoff and the other neighbors' total payoff. Subsequently, agent $j$ reserves the neighboring $k$'s strategy at time step $t$ to the policy at time step $t+1$ with a probability of 
\begin{equation}
	P(k^{t} \to j^{t+1}) = \frac{1}{1+\exp[\frac{(P_j^{t}-P_k^{t})(1-I\gamma)}{\kappa}]}\,.
\end{equation}
    Otherwise, agents incorporate their own chosen strategy from the simulated game at time $t$ into their policy. In this equation, $I$ represents the number of iterations of the simulated game, $\gamma$ is a decay factor and $\kappa$ represents the noise factor, both of which characterize the level of future irrationality in decision making. Note that $\gamma$ can directly determine the size of the node policy with size equals to $1/(1 - 10\gamma)$, and set $\gamma \in [0, 0.1)$. In our present work, the maximum size of the policy is fixed as $10$. 
	  In subsequent rounds of real games, agents follow the strategies determined in the simulated game sequentially until the policy is depleted, followed by new rounds of simulated games to determine a new policy.

    In the following, we employ a series of mathematical definitions to elucidate this model.
    First, we assume that the strategy evolution exhibits a Markov property, meaning that an agent's strategy choice at any given time depends solely on the previous moment, with the strategy choice probability determined by the payoffs, the level of irrationality in decision making, the actual strategies of the agents and their neighbors at the prior time step. Accordingly, the Markov chain is non-homogeneous.

    Existing stochastic theory and dynamic analysis provide us with theoretical guidance. Following established replicator dynamics and Markov dynamics theory, we offer theoretical insights into existing models.
    
	\begin{Definition}     \label{Def_1}
		For a strategy sequence \{$X_t = \xi_t, t \ge 0$\}, we define it as a Markov process on the finite state space $\Theta$ and $\xi_t \in \Theta$ for any $t \ge 0$ and $\xi_t$, the following expression holds:
	\begin{equation}
		\begin{aligned}    
		  &P\{X_{t+1}=\xi_{t+1}|X_t=\xi_t,\dots, X_0 =\xi_0\} \\
		  &=P\{X_{t+1}=\xi_{t+1}|X_t=\xi_t\} \\   
		\end{aligned}\,.
	\end{equation}
		\end{Definition}
        
	\begin{Definition}     \label{Def_2}
        Let $Q(t)$ be the one-step transition matrix of the Markov chain $\{X_t = \xi_t, t \ge 0\}$. Denote $q_{\xi_i \xi_j}(t)$ as the one-step transition probability from strategy $\xi_i$ to strategy $\xi_j$.
	\end{Definition}
	
    We assume that the agent's game states follow a Markov decision process, with the finite state space denoted as $ G = \{ G_1, \ldots, G_n \} $. Furthermore, it possesses two transition matrices $P$ and$ P^\prime$, which we generalize here to $P^{\xi_i}$, where $ \xi_i \in \Theta $. Next, we consider a sequence \{($\xi_t, g_t), t \ge 0$ \}, where $\xi_t \in \Theta$ and $g_t \in G$. The sequence $\{(\xi_t, g_t), t \ge 0\}$ is a joint stochastic process with finite state space $\Theta \times G$ that satisfies:
		\begin{equation}
			\begin{aligned}
				&P\{X_{t+1}=(\xi_{t+1}, g_{t+1})|X_t=(\xi_t, g_t)\} \\
                &= P_{(\xi_t, g_t)\to(\xi_{t+1}, g_{t+1})}^t\,.
			\end{aligned}
		\end{equation}
        
    The model we have established is a two-dimensional joint stochastic process composed of two Markov processes and it exists in the form of a random vector at each time step. We can further derive the one-step transition probabilities of this joint process without proof as follows:
	\begin{equation}
		\begin{aligned}
		P_{(g_i,\xi_i)\to(g_j,\xi_j)}^t=q_{\xi_i\xi_j}(t)p^{\xi_j}_{g_ig_j}\,.
		\end{aligned}
		\label{eq:5}
	\end{equation}

    We established a two-dimensional joint stochastic process and provided the one-step transition probabilities Eq.~\eqref{eq:5} for this joint process. To better characterize the time evolution of states, we further derive the recursive relationship. Specifically, we focus on how the proportions of different states and strategies evolve over time within this non-homogeneous joint Markov chain. Based on this, we present the following propositions, which provides the mathematical expression for this recursive relationship.
    \begin{proposition}     \label{Theorem1}
        The recurrence relation of the inhomogeneous joint Markov chain is expressed as follows:
        \begin{equation}
		\begin{aligned}
            \rho_{g_i}(t+1)= \sum_{g_j}\sum_{\xi_i}\sum_{\xi_k}\rho_{g_j}(t)\rho_{\xi_i}(t)q_{\xi_i\xi_k}(t)p^{\xi_k}_{g_jg_i}\,,
		\end{aligned}
	\end{equation}        
        where $\rho_{g_i}(t)$ represents the proportion of agents in state $g_i$ at time $t$, while $\rho_{\xi_i}(t)$ denotes the proportion of agents adopting strategy $\xi_i$ at time $t$.
    \end{proposition}
    
    \begin{Proof}
        From the law of total probability, we have:
        \begin{equation}
		\begin{aligned}
		\rho_{g_i}(t+1) &= \sum_{g_j}\rho_{g_j}(t)P_{(g_j\to g_i)}(t) \\
        &=\sum_{g_i}\sum_{\xi_i}\rho_{g_j}(t)\rho_{\xi_i}(t)P^{\xi_i}_{(g_j \to g_i)}(t)\\
        &=\sum_{g_j}\sum_{\xi_i}\sum_{\xi_k}\rho_{g_j}(t)\rho_{\xi_i}(t)P_{(g_j,\xi_i)\to(g_i,\xi_k)}^t    \\
        &=\sum_{g_j}\sum_{\xi_i}\sum_{\xi_k}\rho_{g_j}(t)\rho_{\xi_i}(t)q_{\xi_i\xi_k}(t)p^{\xi_k}_{g_jg_i} \\
		\end{aligned}
	\end{equation}
        Results follow. $\hfill\blacksquare$   
    \end{Proof}
    Proposition~\ref{Theorem1} presents the overall iterative equation of our model, which indicates that the proportion of a given state in the network at a specific time is influenced by the distribution of states and strategies across the network at that time, as well as by the transition probability of states and strategies. This proposition provides a foundation for conducting targeted simulations to explore the consequences of these factors.
    
        \begin{proposition}     \label{Theorem2}
        The stationary distribution of each game state satisfies:
    \begin{equation}
		\begin{aligned}
		\rho_{g_i}=\sum_{\xi_i}\rho_{\xi_i}\pi_{g_i}^{\xi_i}\,,
		\end{aligned}
	\end{equation}
        where $\rho_{g_i}$ represents the stable proportion of agents in state $g_i$, $\rho_{\xi_i}$ denotes the stable proportion of agents adopting strategy $\xi_i$, and $\pi^{\xi_i}_{g_i}$ refers to the limiting probability of state $g_i$ in the transition probability matrix driven by strategy $\xi_i$.
    \end{proposition}
    \begin{Proof}   \label{Proof2}
        Known from the conditional probability that:
        \begin{equation}
		\begin{aligned}
		\rho_{g_i}(t)=\sum_{\xi_i}\rho_{\xi_i}(t)\rho_{g_i}^{\xi_i}(t)\,.    
		\end{aligned}
	\end{equation}
        By applying limits to both sides of the equation, we obtain:
        
        \begin{equation}
		\begin{aligned}
		\lim_{t\to\infty}\rho_{g_i}(t)&=\lim_{t\to\infty}\sum_{\xi_i}\rho_{\xi_i}
        (t)\rho_{g_i}^{\xi_i}(t)    \\
        \rho_{g_i}&=\sum_{\xi_i}\rho_{\xi_i}\pi_{g_i}^{\xi_i}
		\end{aligned}
        \label{eq:lim}
	\end{equation}  
        Results follow. $\hfill\blacksquare$
    \end{Proof}
    
    In Eq.~(\ref{eq:lim}), we assume the existence of a limiting probability $\pi_{g_i}^{\xi_i}$. A sufficient condition for this assumption is as follows: the transition probability matrix driven by strategy $\xi_i$ is irreducible and positive recurrent in all states. Furthermore, we assume that the operations of taking the limit and summation can be interchanged, which is valid for most cases, including birth-death processes and all finite-state models, such as our current model.    
    Proposition~\ref{Theorem2} indicates that, as time approaches infinity, the frequency of each game state corresponds to the limiting distribution of the transition probabilities driven by each strategy, aggregated according to the proportions of strategies.
   
	\section{Simulation and Analysis}		\label{Simu}

    Next, we explore the effectiveness of the aforementioned model through a series of simulations. First, in Sec.~\ref{Methods}, we introduce the detailed configuration of the networked population and game models used for these simulations. Next, in Sec.~\ref{TG} and~\ref{GG}, we analyze how the parameter settings of the Markov decision chain and the payoff matrix parameters of the game models influence the proportion of different game states in the population. At the same time, we validate the previously mentioned theories and calculate the simulation errors. We then conduct simulations to verify the proposed proposition. Finally, we discuss the specific impact of these parameters on the cooperation level in Sec.~\ref{TPC} and in Sec.~\ref{GC}.
 
	\subsection{Methods}	\label{Methods}
    We first provide a detailed explanation of the parameter settings for the game model and the networked population used in the following simulations.
	  As a key ingredient of our model, agents are allowed to engage in games with different states, hence we introduce three variants of the donation game model by using different parameter values for each case. Specifically, based on the definitions in Sec.~\ref{Model}, the payoff parameters for $G_i$ are set as $R_i = b_i - c$, $S_i = -c$, $T_i = b_i$, and $P_i = 0$. Here $c$ represents the cost incurred when an agent chooses to cooperate, while $b_i\ge c$ denotes the benefit an agent receives when the neighbor cooperates. We assume that $b_1 \le b_2 \le b_3$, and $b_1+2\Delta = b_2 + \Delta = b_3$ meaning that while the cost of cooperation $c$ is the same across the three game models, the benefit of cooperation increases progressively. Thus, the highest benefit is in $G_3$, while it decreases in $G_2$ and $G_1$, but the difference $\Delta$ between various games is fixed. In agreement with several previous works, we set $\kappa=0.1$ for the noise parameter. Since the only Nash equilibrium in the donation game is mutual defection, where the payoff from mutual defection ($P_i=0$) is significantly lower than that from mutual cooperation ($R_i=b_i-c$), our model reflects a typical social dilemma faithfully. In the initial state, the strategy profile is randomly selected from $\{C, D\}$ set for all agents, and the game states are also randomly generated from $\{G_1, G_2, G_3\}$ set.
 
    Next, regarding the Markov decision chain for state transitions of agents, we define two game transition probability matrices, 
\begin{equation}
\begin{aligned}
	P=\left(\begin{array}{lll}
    	\mu - \delta & \sigma + \delta  & 0 \\
    	\mu	- \delta& 1-(\mu +\sigma) & \sigma + \delta \\
    	0	& \mu - \delta & \sigma + \delta \\
\end{array}\right),\,   \\
	P^\prime=\left(\begin{array}{lll}
			\mu + \delta&\sigma - \delta  & 0\\
			\mu	+ \delta& 1-(\mu +\sigma) & \sigma - \delta\\
			0	& \mu + \delta &\sigma - \delta \\
\end{array}\right)\,,
\end{aligned}
\end{equation}
    where matrix $P$ describes the transition of an agent’s game state at the next time step when they choose to cooperate, while matrix $P^\prime$ describes the transition when they choose to defect. The parameters $\sigma$ and $\mu$ in these matrices control the probability of transitioning to $G_1$ and $G_3$ respectively. It means that agents will change the game state to $G_3$ more easily as $\sigma$ increases, while agents will easier to change the game state to $G_1$ as $\mu$ increases. We set $\sigma + \mu =1$ to ensure the validity of the probability matrix, and the parameter $\delta$ controls the difference between $P$ and $P^\prime$. It is important to note that as $\delta$ increases, agents who defect are more likely to transition to $G_1$, whereas cooperation tends to maintain the game state of agents in $G_3$ more easily.
    In the simulations setup, we conduct $10^3$ rounds of games on a WS network with $3000$ nodes where $k=6$ and $p=0.4$ to ensure that the game states and cooperation frequency reach convergence. All simulation results are averaged over five independent runs. 

	\subsection{Effect of Transition Parameters on the Proportion of Different Game States}		\label{TG}
 
\begin{figure}
	\centering
	\begin{subfigure}[h]{0.48\linewidth}   
		\centering
		\includegraphics[width=1\linewidth]{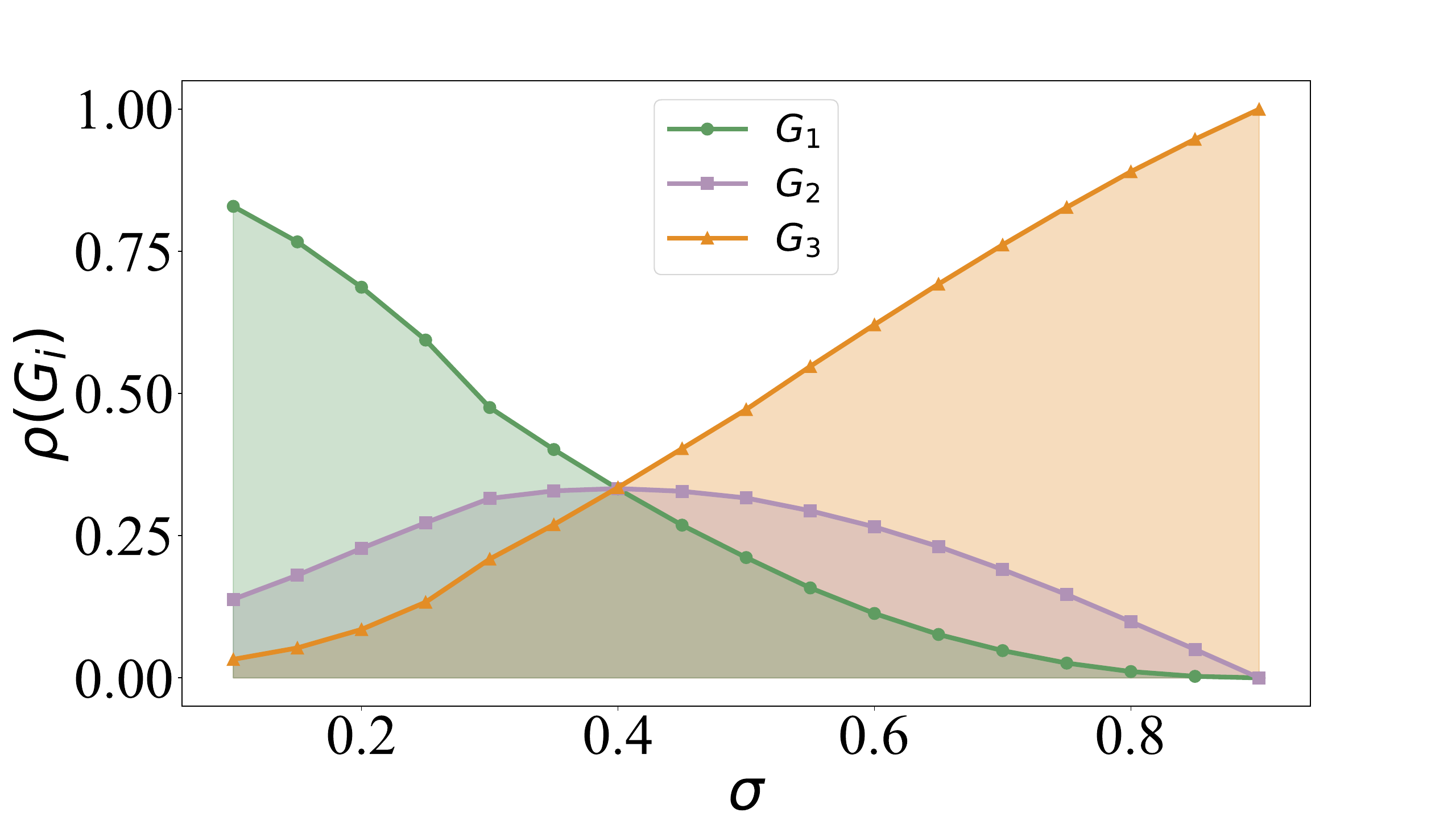}
            \caption{Effect of $\sigma$}
		\label{fig2a}
	\end{subfigure}
	\begin{subfigure}[h]{0.51\linewidth}   
		\centering
	\includegraphics[width=1\linewidth]{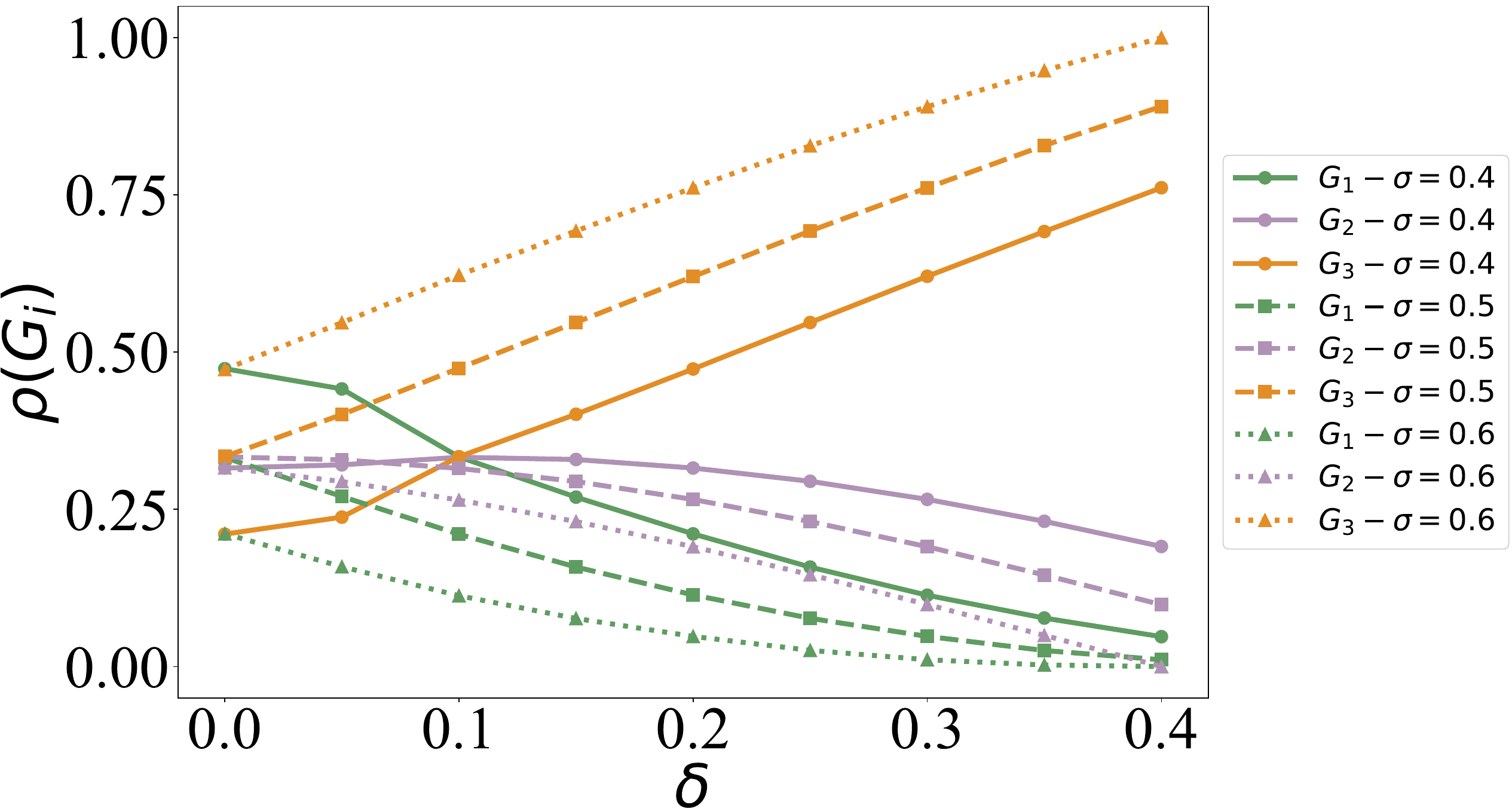}
        \caption{Effect of $\delta$}
        \label{fig2b}
    \end{subfigure}
            \caption{\textbf{The effect of $\sigma$ and $\delta$ on the proportion of different game states.} In (\subref{fig2a}), circles, squares, and triangles represent the density of three game states ${G_1, G_2, G_3}$ as $\sigma$ increases. The other parameters are $\delta = 0.1, \gamma=0 , b_1 = 3, \Delta = 1$ and $c = 1$. In (\subref{fig2b}), solid lines with circles, dashed lines with squares, and dotted lines with triangles correspond to $\sigma$ values of $0.4$, $0.5$, and $0.6$, respectively. According to the legend, different colors represent the densities for game states ${G_1, G_2, G_3}$ as $\delta$ varies. The other parameters are $b_1 = 3$, $\Delta = 1$, and $c = 1$. The simulation results are obtained after $10^3$ iterations on the WS network.}
 	\label{fig:3}
    
\end{figure}

    Before focusing on the distribution of cooperators within the population, it is crucial to analyze the density of each game state. To address this, our simulations investigate three game states ${G_1, G_2, G_3}$ introduced in Sec.~\ref{Methods} and examine how their densities vary with the parameters of the Markov decision chain, $\sigma$, and $\mu$. Given that $\sigma + \mu = 1$, our analysis will focus on the impact of $\sigma$ on the density of each game state, with $\mu$ as a complement. The results are displayed in Fig.~\ref{fig:3}.
	
    Fig.~\ref{fig2a} presents the results of each game state in the WS network with different values of the transition parameter $\sigma$. Note that at $\gamma=0$ there is no difference from the traditional imitation method. A clear trend can be observed: as $\sigma$ increases, the density of $G_3$ in the population initially increases, with an intersection point occurring at $\sigma^* + \delta = 0.5$, where $\sigma^*=0.4$. In contrast, the density of $G_1$ exhibits the opposite trend and decays with with $\sigma$. Its value agrees with $G_3$ at $\sigma = \sigma^*$. Interestingly, the density of $G_2$ exhibits a non-monotonous trend, first up, later down, and reaching a maximum value at $\sigma=\sigma^*$. 
    This indicates that the system gradually shifts from game state $G_1$ to $G_2$ when $\sigma < \sigma^*$. Beyond this threshold the transition from game state $G_2$ to $G_3$ becomes relevant. When $\sigma$ is sufficiently large then the density of $G_3$ in the population approaches $1$, indicating that the population exclusively occupies the $G_3$ state without using alternative game states. This feature is restricted to $G_3$ and is not observed for the other two game states.
    
\begin{figure}[htbp]
	\centering
	\begin{subfigure}[t]{0.25\linewidth}   
		\centering
		\includegraphics[width=\linewidth]{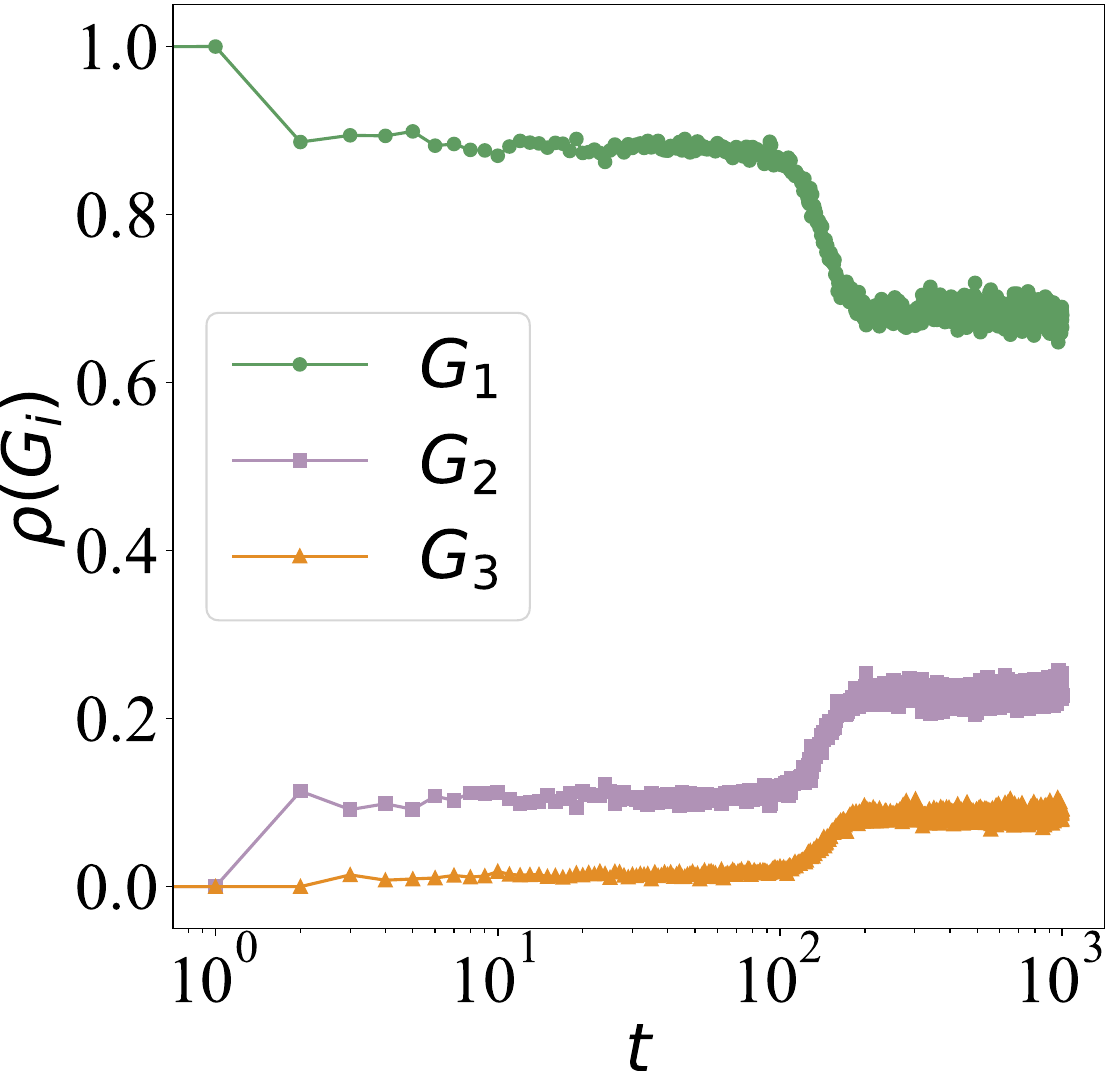}
		\caption{$\sigma=0.2$}
            \label{figsig:a}
	\end{subfigure}
	\begin{subfigure}[t]{0.25\linewidth}    
		\centering
		\includegraphics[width=\linewidth]{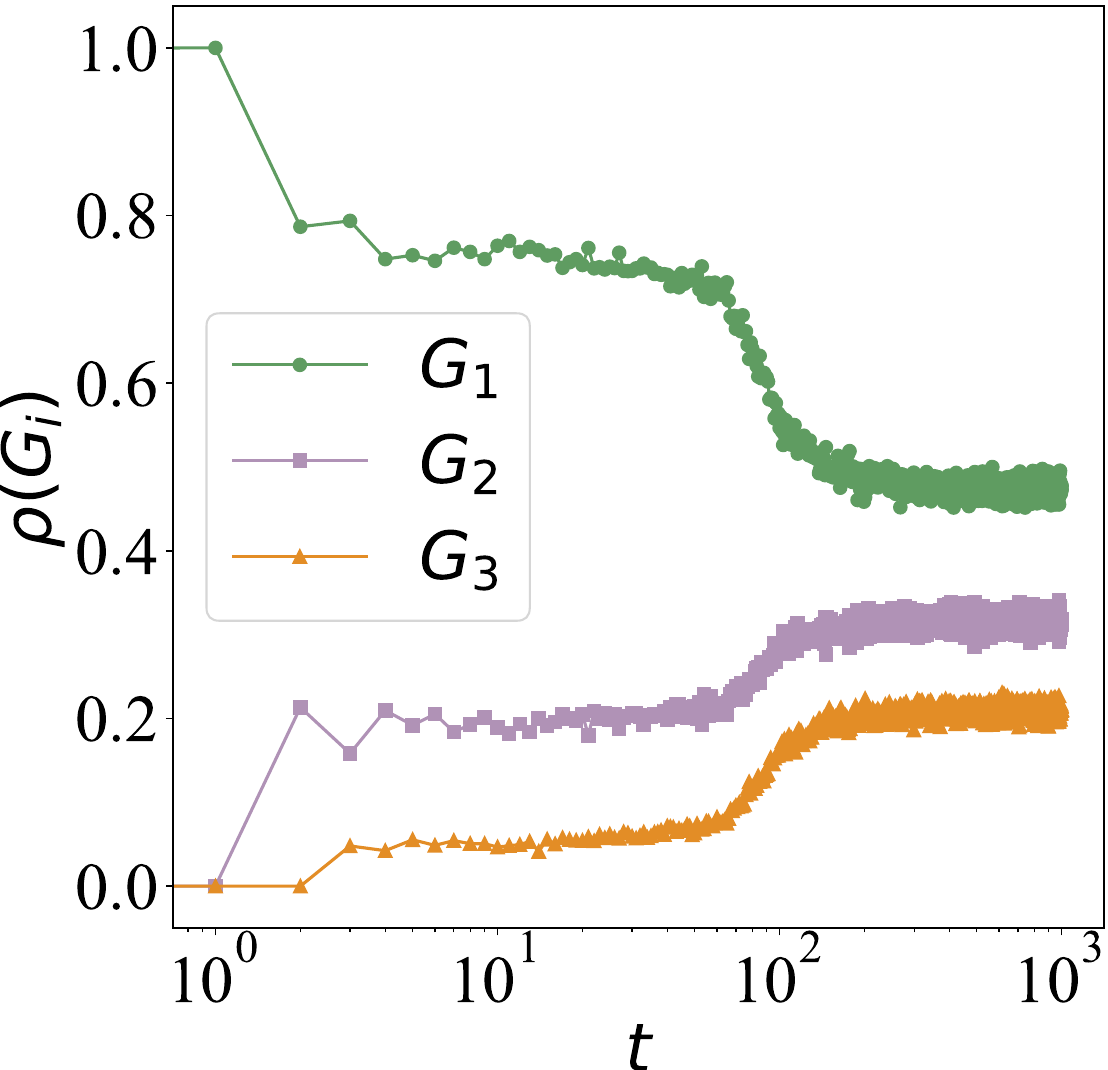}
		\caption{$\sigma=0.3$}
            \label{figsig:b}
	\end{subfigure}
	\begin{subfigure}[t]{0.25\linewidth}    
		\centering
		\includegraphics[width=\linewidth]{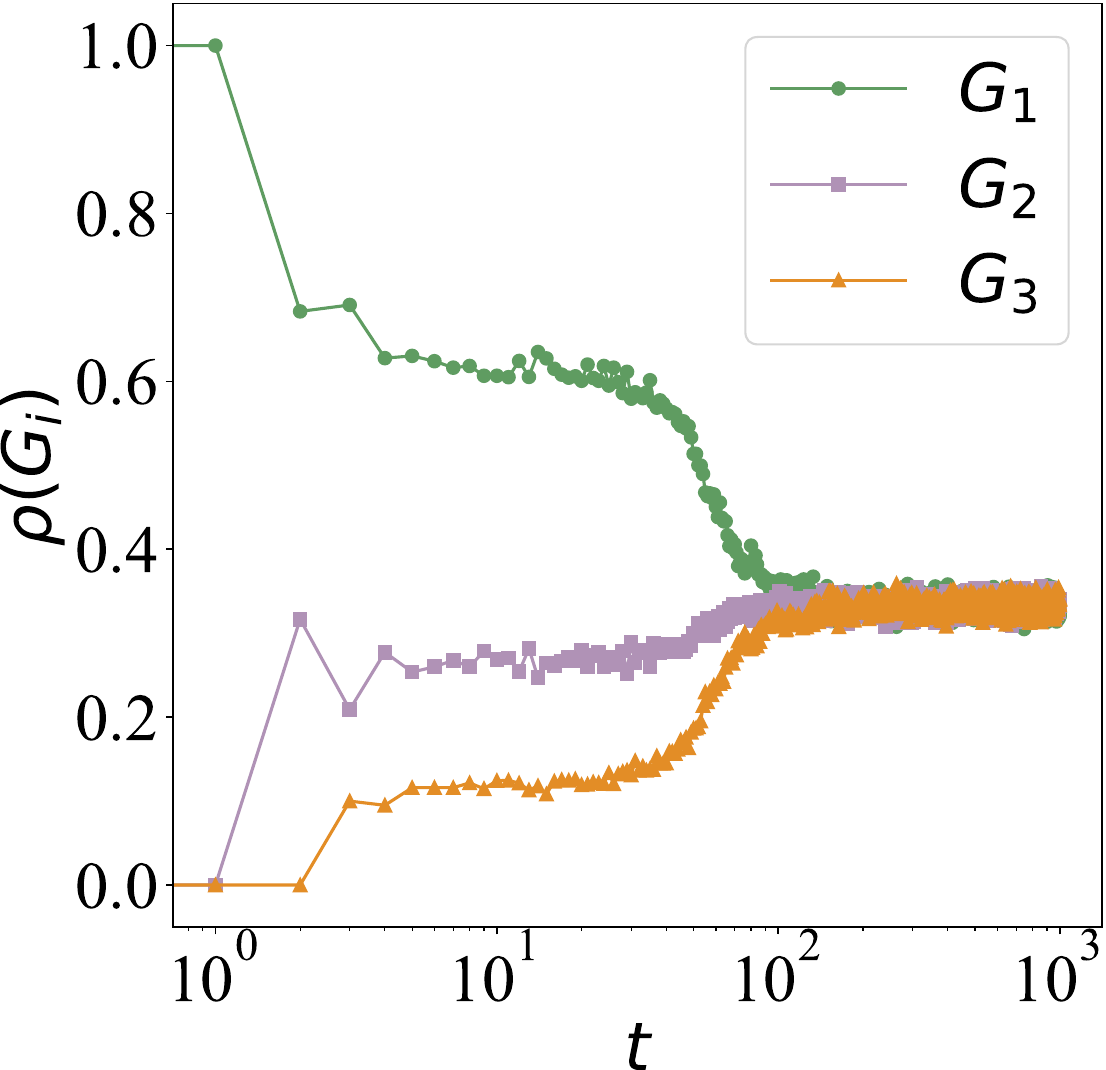}
		\caption{$\sigma=0.4$}
            \label{figsig:c}
	\end{subfigure}
    
	\begin{subfigure}[t]{0.25\linewidth}    
		\centering
		\includegraphics[width=\linewidth]{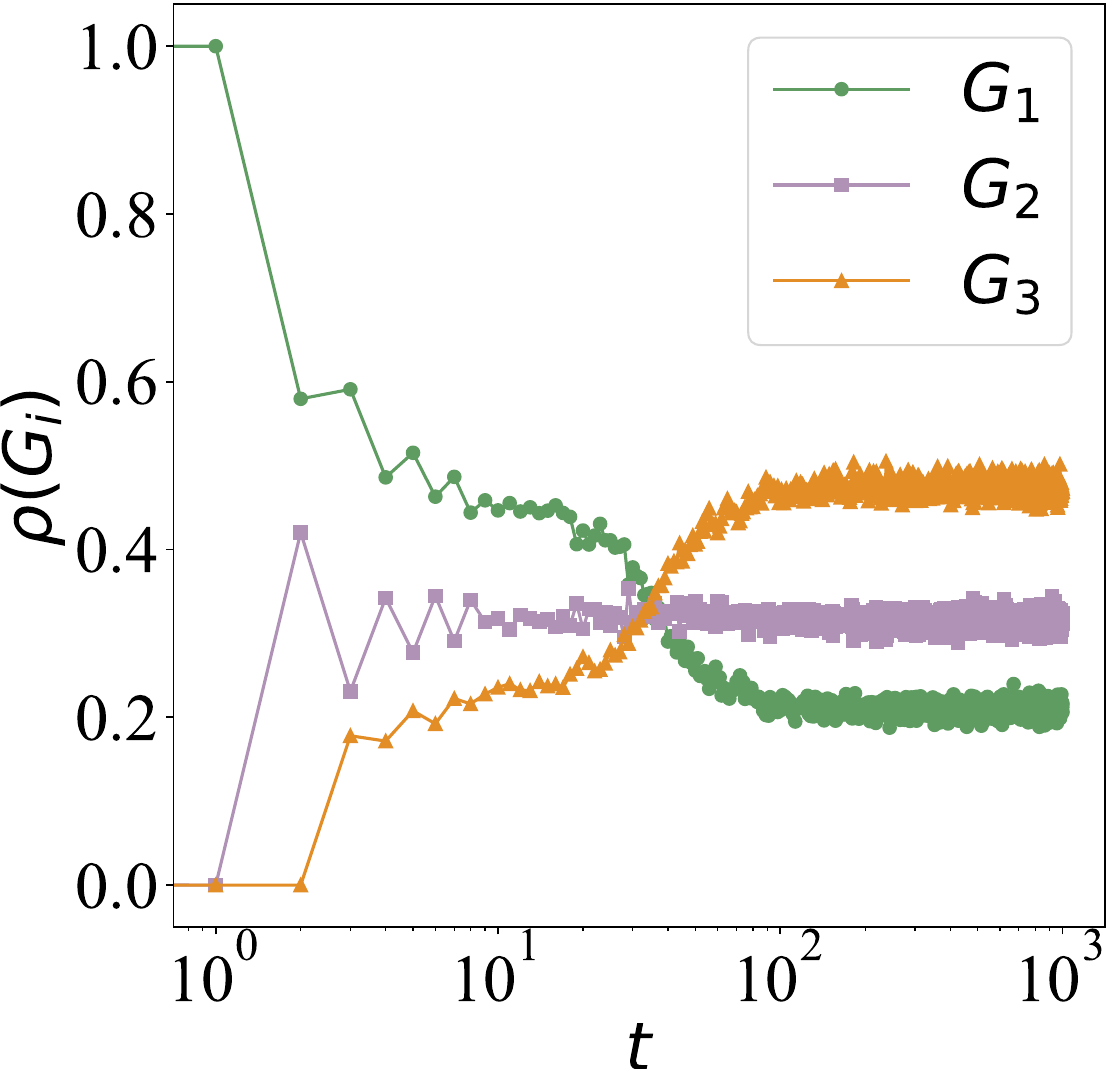}
		\caption{$\sigma=0.5$}
            \label{figsig:d}
	\end{subfigure}
	\begin{subfigure}[t]{0.255\linewidth}    
		\centering
		\includegraphics[width=\linewidth]{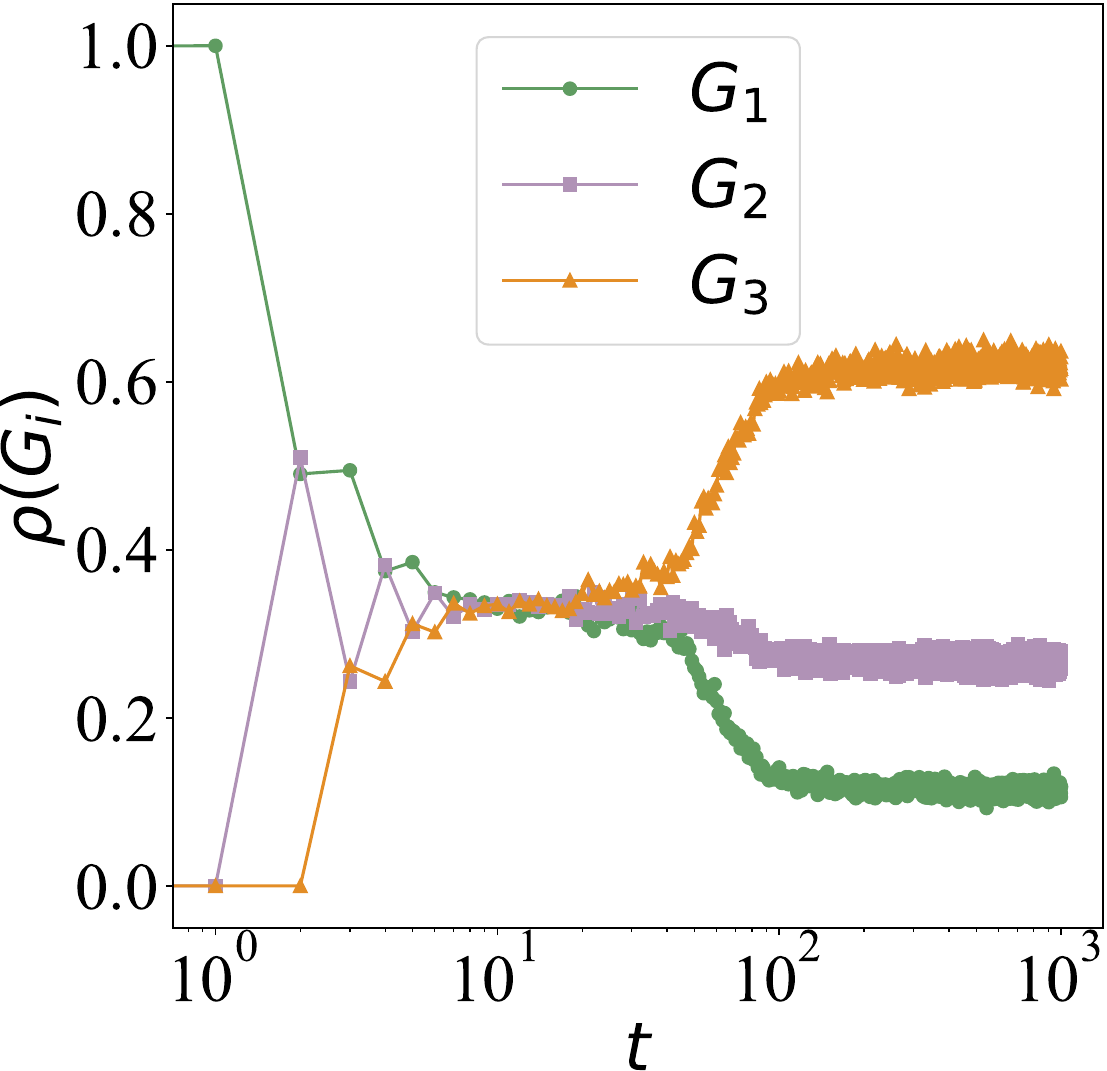}
		\caption{$\sigma=0.6$}
            \label{figsig:e}
	\end{subfigure}
	\caption{\textbf{The impact of $\sigma$ on the evolution of different game states}. Panels show the time evolution of the density of game states ${G_1, G_2, G_3}$ at different $\sigma$ values, as indicated. Parameters are $\delta = 0.1$, $b_1 = 3$, $\Delta = 1$, and $c = 1$.}
	\label{fig3}
\end{figure}

    Next, we examine how $\delta$ influences the dynamics within each game state. The results are summarized in Fig.~\ref{fig2b}. Our findings reveal that the density of $G_3$ gradually rises as $\delta$ increases, while the densities of the other two game states decline. Furthermore, we identify a critical point $\delta^*$, where $\delta + \sigma = 0.5$. When $\delta = \delta^*$, the densities of all game states converge to the same value of $1/3$. When $\delta > \delta^* $, the density of $G_3$ exceeds that of $G_1$, indicating a game state resistant to improvement through cooperative behavior. Conversely, when $\delta < \delta^*$, the density of $G_1$ surpasses that of $G_3$, suggesting an improvement of the game state through cooperative behavior.
    
    To further investigate how parameter $\sigma$ influences the observed patterns of game states, we examine five specific values near $\sigma^*$. Fig.~\ref{fig3} displays the temporal density dynamics for game states ${G_1, G_2, G_3}$ at $\sigma$ values of $0.2, 0.3, 0.4, 0.5$, and $ 0.6$. For a proper comparison, we apply an alternative initial state where all agents are launched from the game state $G_1$.
    As $\sigma$ increases, the convergence patterns across different game states align with those observed in previous simulations. Furthermore, the density of $G_3$  reaches the same level as the other game states. The relaxation time at which the density of $G_3$ begins to saturate for various $\sigma$ values is nearly identical. This phenomenon can be attributed to the fact that, although the density of $G_3$ is comparable to the other game states in the early stages, the cooperation frequency within the population has not been able to increase rapidly in response to improvements in the game states. This results in a form of system's inertia. This system's inertia phenomenon is most evident in Fig.~\ref{figsig:e}. Consequently, while the density of $ G_3 $ can quickly reach a level comparable to other game states, further convergence does not occur immediately.
	 
    Finally, we compare the simulation results with the proposition previously discussed by setting $\sigma = 0.3$, $b_1 = 3$, $\Delta = 1$, and $\gamma = 0$. As shown in Table~\ref{simu:theorem}, the relative errors between the results are within 1\%, demonstrating our proposition's validity and indirectly confirming the robustness of the model.
    
\begin{table*}[htbp]
    \centering
    \caption{Stationary Portion of Agents Staying in Different Game States}
    \begin{tabular}{c|ccc|ccc}
    \hline 
    \multirow{2}{*}{Results} & \multicolumn{3}{c|}{$\delta=0.02$} & \multicolumn{3}{c}{$\delta=0.04$} \\
    \cline{2-7} 
    & $G_1$ & $G_2$ & $G_3$ & $G_1$ & $G_2$ & $G_3$ \\
    \hline 
    Simulation results & 0.4671 & 0.3172 & 0.2157 & 0.4493 & 0.3204 & 0.2302 \\
    Theoretical results & 0.4681 & 0.3165 & 0.2154 & 0.4500 & 0.3189 & 0.2310 \\
    Relative error (\%) & 0.214 & 0.221 & 0.139 & 0.156 & 0.468 & 0.348 \\
    \hline
    \end{tabular}
    \vspace*{-4pt}
    \label{simu:theorem}
\end{table*}
    
    Notably, although in the transition matrices $ P^\prime$, the probability for defectors to retain the game states in $G_1$ increases with a higher $\delta$, the population density of $G_1$ still demonstrates a downward trend. 

\subsection{Effect of Game Parameters on the Proportion of Different Game States}		\label{GG}

\begin{figure}
	\centering
	
	\begin{subfigure}[t]{0.3\linewidth}
		\centering
		\includegraphics[width=\linewidth]{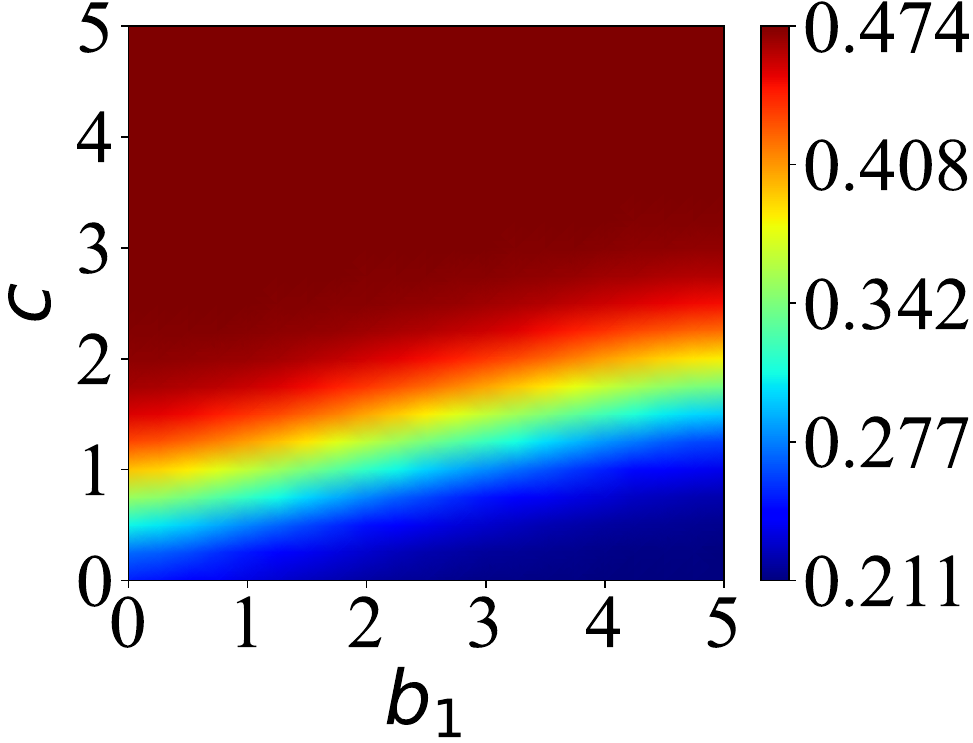}
		\caption{Effect of $b_1$ and $c$ on $G_1$}
            \label{G1:A}
	\end{subfigure}
  	\begin{subfigure}[t]{0.3\linewidth}
  		\centering
  		\includegraphics[width=\linewidth]{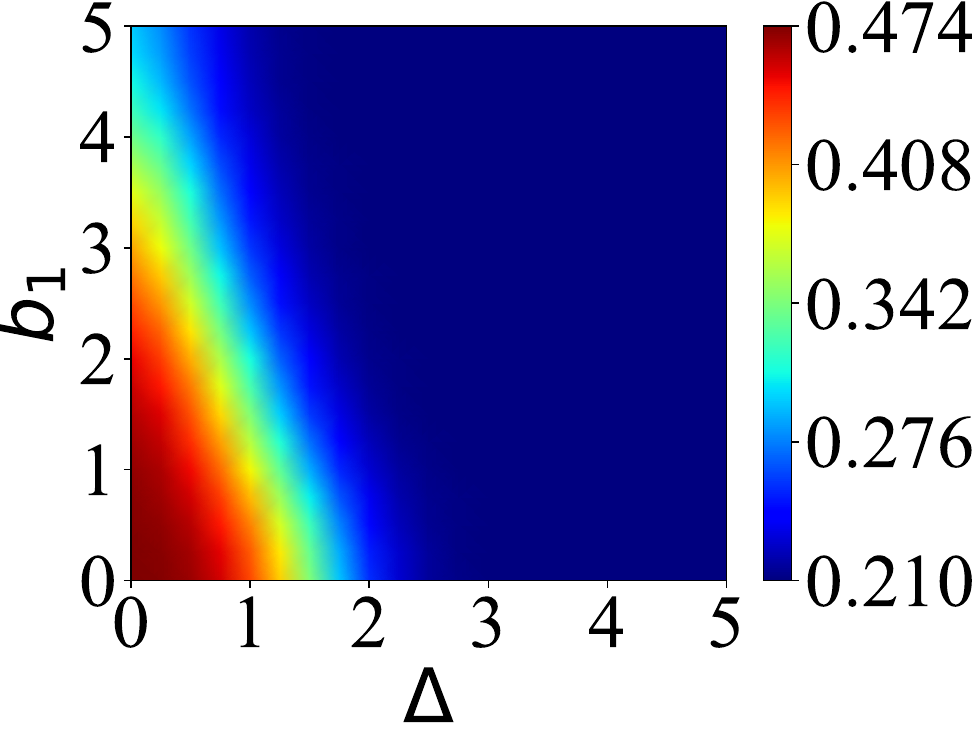}
  		\caption{Effect of $b_1$ and $\Delta$ on $G_1$}
            \label{G1:B}
  	\end{subfigure}
	\begin{subfigure}[t]{0.3\linewidth}
		\centering
		\includegraphics[width=\linewidth]{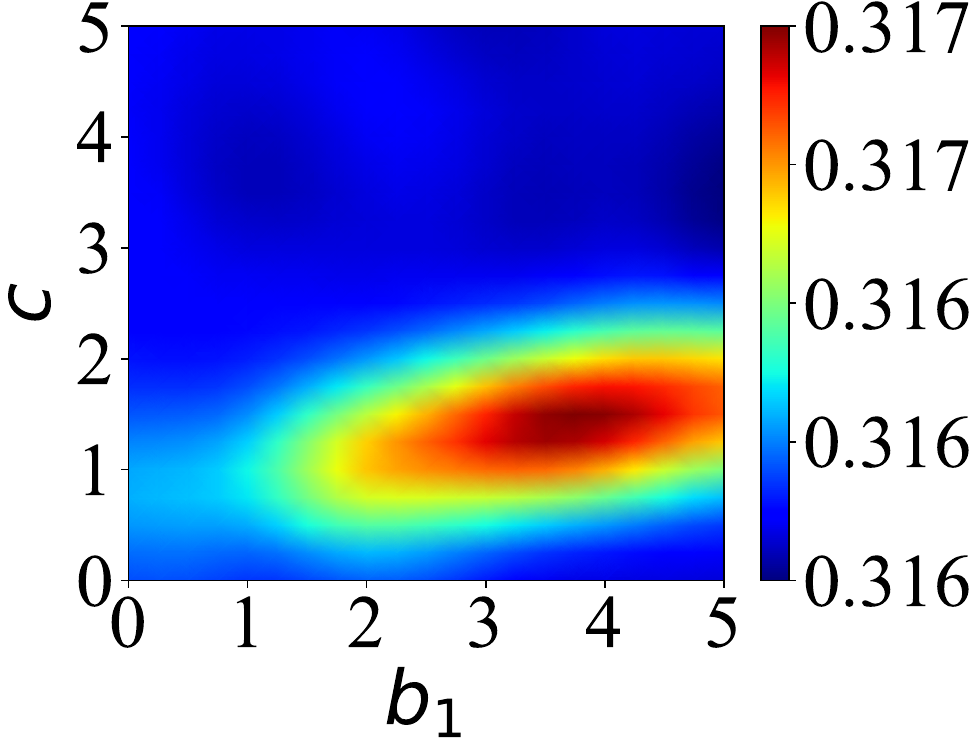}
		\caption{Effect of $b_1$ and $c$ on $G_2$}
            \label{G2:A}
	\end{subfigure}
  	\begin{subfigure}[t]{0.3\linewidth}
  		\centering
  		\includegraphics[width=\linewidth]{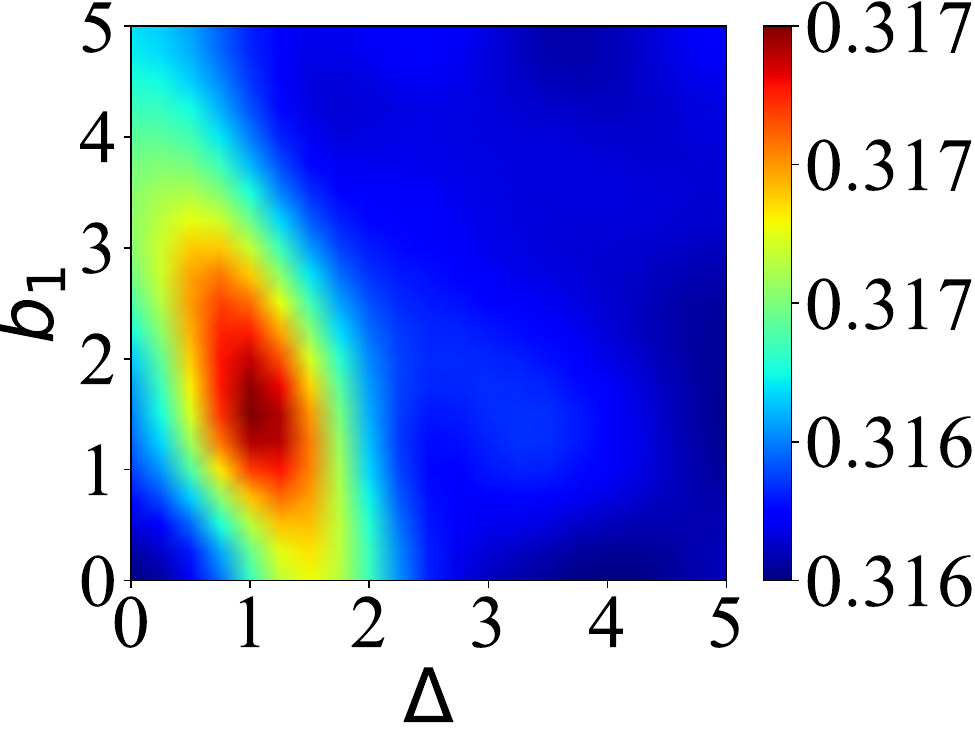}
  		\caption{Effect of $b_1$ and $\Delta$ on $G_2$}
            \label{G2:B}
  	\end{subfigure}
  	\begin{subfigure}[t]{0.3\linewidth}
  		\centering
  		\includegraphics[width=\linewidth]{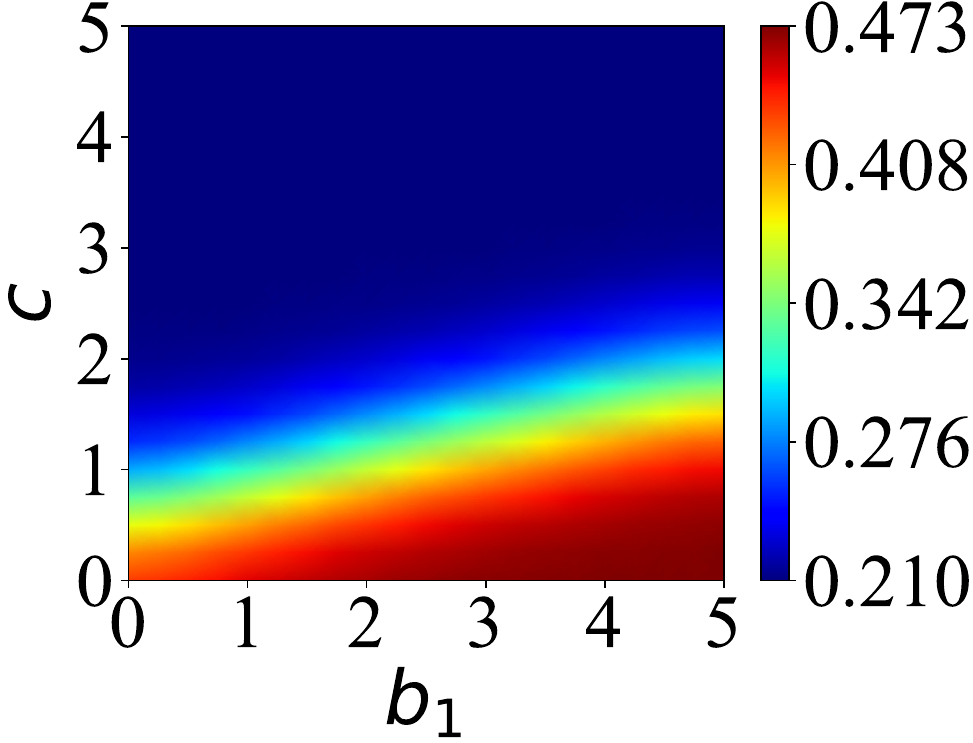}
  		\caption{Effect of $b_1$ and $c$ on $G_3$}
            \label{G3:A}
  	\end{subfigure}
  	\begin{subfigure}[t]{0.3\linewidth}
  		\centering
  		\includegraphics[width=\linewidth]{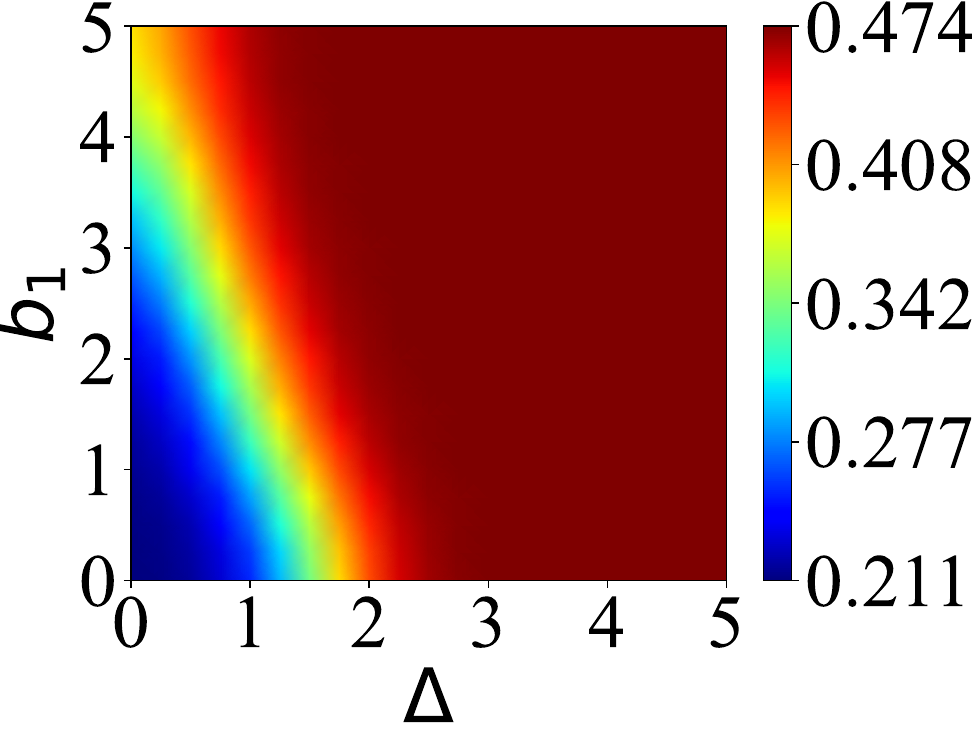}
  		\caption{Effect of $b_1$ and $\Delta$ on $G_3$}
            \label{G3:B}
  	\end{subfigure}
    
        \caption[]{\textbf{The effect of game parameters on the proportion of game states}. (\subref{G1:A}): we fix $\Delta=1$ to explore the effect of $c$ and $b_1$ on the density of $G_1$. (\subref{G1:B}): we fix $c=1$ to explore the effect of $b_1$ and $\Delta$ on the density of $G_1$. The other parameters are $\sigma = 0.5$, $\delta = 0.1$, and $\gamma= 0$. For the effect of game parameters on the proportion of $G_2$, we set $\sigma = 0.5$, $\delta = 0.1$, and $\gamma= 0$. Furthermore, $\Delta=1$ in panel~(\subref{G2:A}), while $c=1$ in panel~(\subref{G2:B}). Note that there is only a tiny difference between the minimal and maximal values in these panels. For the effect of game parameters on the proportion of $G_3$ we use $\sigma = 0.5$, $\delta = 0.1$, and $\gamma= 0$. In panel~(\subref{G3:A}) we fix $\Delta=1$, while in panel~(\subref{G3:B}) $c=1$ is used.}
  		\label{fig:7}
\end{figure}

    We now explore in depth how the payoff matrices of different game states specifically affect the density distribution of each game state, thereby revealing the mechanisms through which the payoff matrices influence the dynamic distribution of the game state. The results are summarized in Fig.~\ref{fig:7}.

    Firstly, we examine the effect on the game state $ G_1$. In this Markov decision chain, the transition probabilities are directly determined by three transition key parameters jointly: $\sigma$, $\mu$, and $\delta$, indirectly determined by three game parameters: $b_1$, $c$, and $\Delta$. The simulation results are shown in Figs.~\ref{G1:A}, \ref{G1:B} by heat maps. In Fig.~\ref{G1:A} we fix $\Delta=1$ and in Fig.~\ref{G1:B} we fix $c=1$ to explore the effect of the other parameters on the density of $G_1$. These panels reveal that as $b_1$, $\Delta$ increases or $c$ decreases, the density of state $G_1$ in the population gradually declines. This is because, with an increase in $b$ or a decrease in $c$, the benefit-to-cost ratio of agents in the group improves, making them more inclined to adopt cooperative strategies. As the proportion of cooperative strategies in the group increases, the overall game state of the population shifts toward $G_2$ and $G_3$ which are more favorable to cooperation, rather than $G_1$. This transition results in an indirect decrease of the density of $G_1$.

    Next, we analyze how the payoff matrix parameters influence the density of game state $G_2$. The simulation results are presented in Figs.~\ref{G2:A}, \ref{G2:B}. In Fig.~\ref{G2:A}, we fix $\Delta=1$, while $c=1$ is fixed in Fig.~\ref{G2:B}. Note that $G_2$ is designed as a transitional state between the extreme states of $G_1$ and $G_3$. Consequently, we find that the density variation of the game state $G_2$ in the population consistently stabilizes around $6/19$ regardless of the values of $b_1$ and $c$. The transition matrices used in our simulation resulted in equal limit frequencies for $G_2$, regardless of whether cooperation is transitioned with the $P$ transition or defection followed by the $P^\prime$ transition. It indicates that cooperation and defection are independent of the density distribution of $G_2$, thereby indirectly suggesting that the payoff matrix of the game does not influence the density distribution of $G_2$. Furthermore, the heat maps we present still exhibit minor fluctuations, which can be attributed to the randomness of initial conditions and the noise associated with strategy evolution during the game process.

    Finally, we present the impact of the payoff matrix elements on the density of the $G_3$ state. In Figs.~\ref{G3:A}, \ref{G3:B} we used the same parameter values applied in Figs.~\ref{G1:A}, \ref{G1:B} and in Figs.~\ref{G2:A}, \ref{G2:B}. The influence of parameters $b_1$ and $c$ on the density of the $G_3$ state is shown in Fig.~\ref{G3:A}. It suggests that as $b_1$ increases or $c$ decreases, the density of the $G_3$ state rises significantly. This is because an increase in $b_1$ or a decrease in $c$ effectively enhances the benefit-to-cost ratio in all game states, leading to an increase in the frequency of cooperators, which in turn indirectly raises the density of the $G_3$ state. In Fig.~\ref{G3:B}, we illustrate the effect of the parameters $b_1$ and $\Delta$ on the density of the $G_3$ state. As the differences of benefit between games increase, the density of the $G_3$ state also increases, further indicating the important role of $\Delta$ in promoting cooperative behavior within the population. It is noted that the effects discussed are complementary to the results observed in $G_1$: as $b_1$ increases or $c$ decreases, agents will leave the $G_1$ state. According to our simulations, the density of $G_2$ remains stable independent of the game parameters. Consequently, it implies that nearly all agents leaving the $G_1$ state will arrive at the $G_3$ state.

  	\subsection{Effect of Transition Parameters and the Size of Policy on the Cooperation Frequency}	\label{TPC}
	
    In the remaining part of this work, we explore how the cooperation level is affected by the parameters characterizing the transition process. Starting with the parameter set by the Markov decision chain, we fix the payoff parameters at $b_1 = 3$, $\Delta = 1$, and $\gamma = 0$, indicating that agents engage in a traditional single-round imitation game. We then examine the trend in the density of cooperators within the population upon convergence as $\sigma$ varies under different values of $\delta$.

\begin{figure}
		\centering
		\begin{subfigure}[t]{0.45\linewidth}
			\centering
			\includegraphics[width=\linewidth]{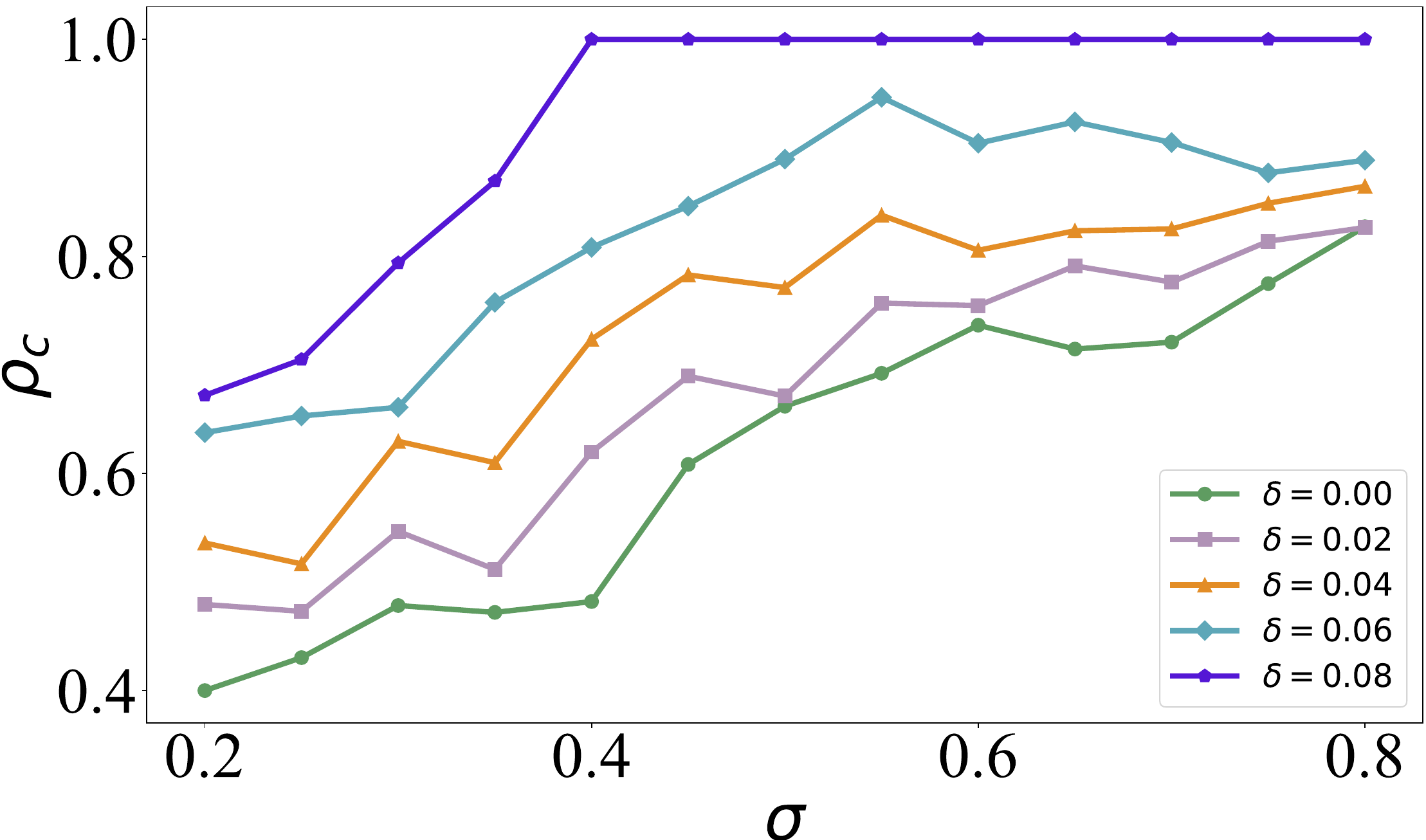}
			\caption{Effect of Transition Parameters}
                \label{fig.5a}
		\end{subfigure}
		\begin{subfigure}[t]{0.45\linewidth}
			\centering
			\includegraphics[width=\linewidth]{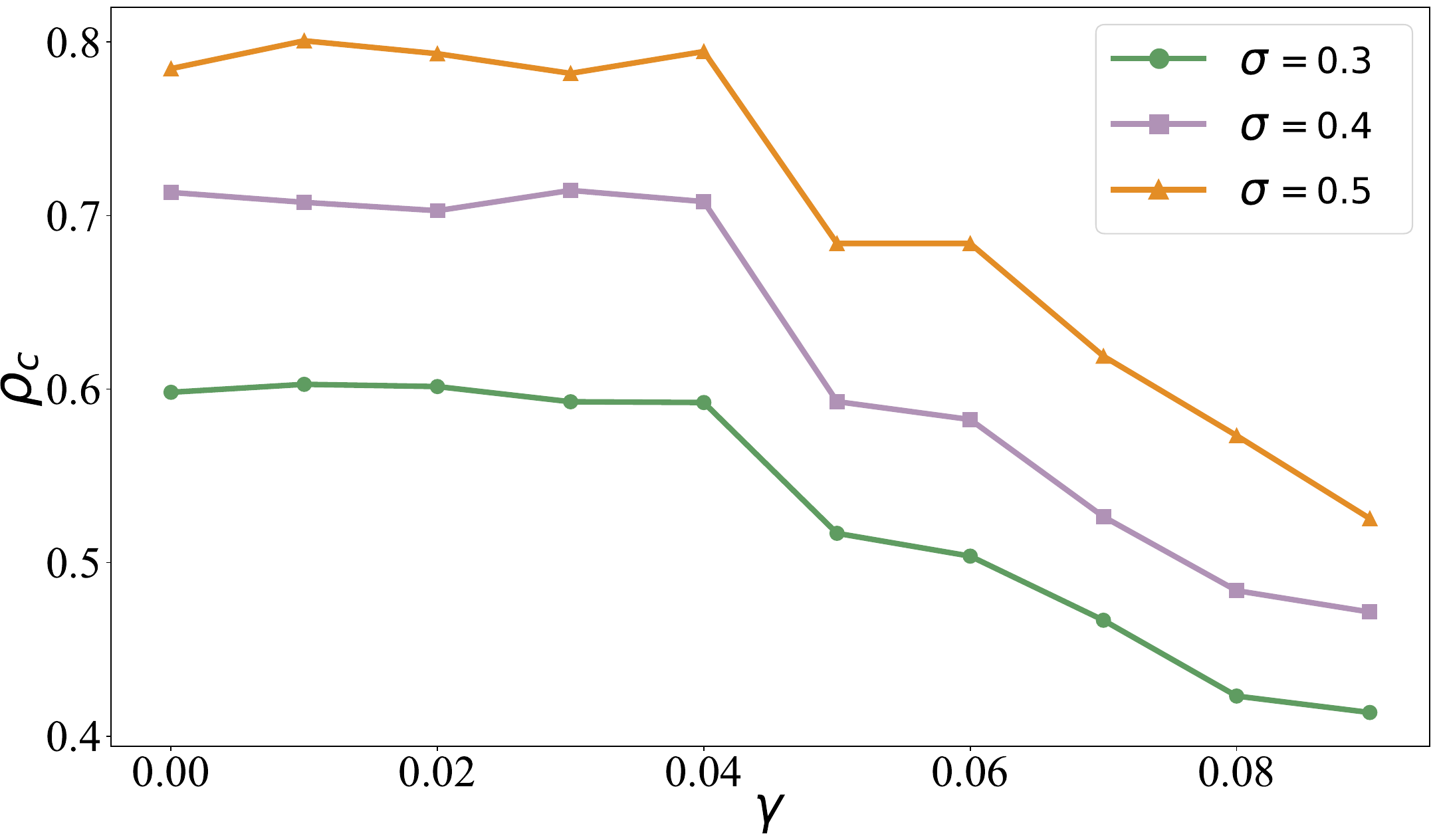}
			\caption{Effect of Policy Size}
                \label{fig.5b}
		\end{subfigure}
                        
            \caption{\textbf{The effect of transition parameters and policy size on the cooperation frequency}. In (\subref{fig.5a}), we set the payoff parameters at $b_1 = 3$, $\Delta = 1$, and $\gamma = 0$. The curves with circular, square, triangular, rhombic, and pentagonal illustrate how the density of Cooperation in the population varies with $\sigma$ for $\delta$ values of $0.00$, $0.02$, $0.04$, $0.06$, and $0.08$. In (\subref{fig.5b}), we select three different values of $\sigma=\{0.3, 0.4, 0.5\}$, represented by curves with circles, squares, and triangles, respectively. The remaining parameters are $\delta=0.04$, $b_1=3$, $c=1$, and $\Delta=1$.}
            \label{fig5}
            \end{figure}

    It is important to note that none of the game states ${G_1, G_2, G_3}$ satisfies the condition $b/c > k$ with the current parameter setting. It implies the clear absence of a cooperation-supporting situation. The simulation results are shown in Fig.~\ref{fig.5a}. The five curves with circular, square, triangular, rhombic, and pentagonal markers illustrate how the density of cooperation in the population varies with $\sigma$ at different $\delta$ values. It is clear that as $\sigma$ and $\delta$ increase, the density of cooperators in the population shows a steady upward trend. This can be explained by the fact that the increase in $\delta$ leads to an increasing correlation between the strategy choices of agents and the changes in the game state. The game state gradually intensifies its pressure on cooperative behaviors, and the increase of $\sigma$ enhances the transition to $G_3$, thus promoting the emergence of cooperation within the population.

    Interestingly, the density of cooperators rises significantly. It is essential to emphasize that, even though the cooperation-dominant condition $b/c > k$ is still unmet at this stage, the influence of defection on the game state transition becomes stronger as $\delta$ increases, making it easier for the system to transition into states unfavorable for cooperation. Nevertheless, despite this unfavorable transition prompted by defection, cooperative behaviors within the population are markedly enhanced, indicating that the game state transition mechanism and the differences between those games play a crucial role in promoting cooperation.

    Next, we analyze the role of the decay factor $\gamma$ which represents the degree of irrationality and the size of the policy on the game states. 
    In Fig.~\ref{fig.5b}, we fix the parameters $\delta=0.04$, $b_1=3$, $c=1$, and $\Delta=1$ and select three different values of $\sigma \in \{0.3, 0.4, 0.5\}$ to illustrate how the density of cooperators changes with $\gamma$. The results indicate that the density of cooperators increases as $\sigma$ increases, consistent with previous experimental findings. Within the range of $\gamma \in (0, 0.05)$, the density of cooperators remains stable. However, when $\gamma$ reaches or exceeds $0.05$, the level of cooperation decreases significantly. This decline is attributed to the expansion of the policy to a second-order size. 
	
	\subsection{Effect of Game Parameters on the Cooperation Frequency}		\label{GC}

\begin{figure}
		\centering
		\begin{subfigure}[t]{0.45\linewidth}
			\centering
			\includegraphics[width=\linewidth]{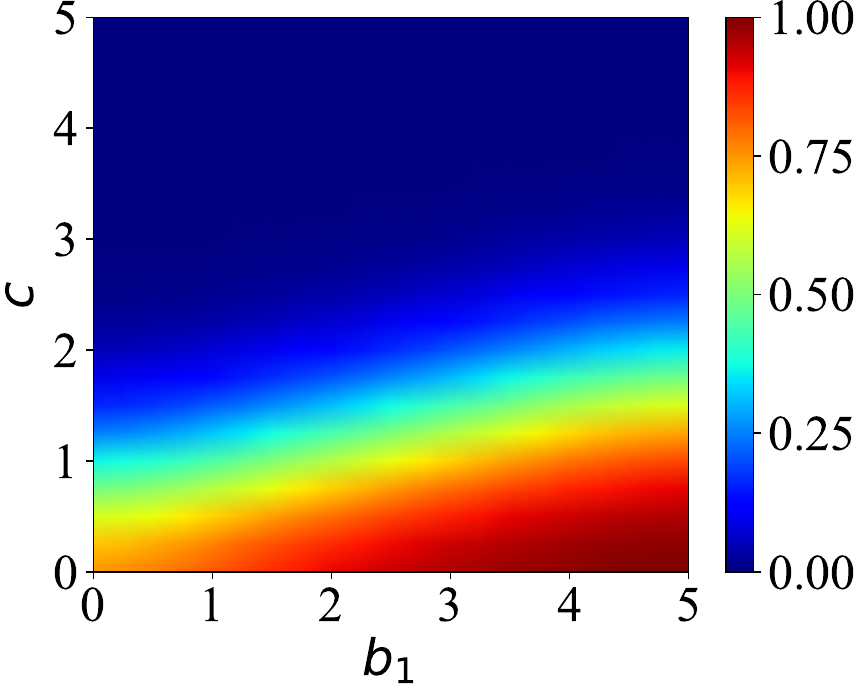}
			\caption{Effect of $b_1$ and $c$ }
			\label{fig6a}
		\end{subfigure}
		\begin{subfigure}[t]{0.45\linewidth}
			\centering
			\includegraphics[width=\linewidth]{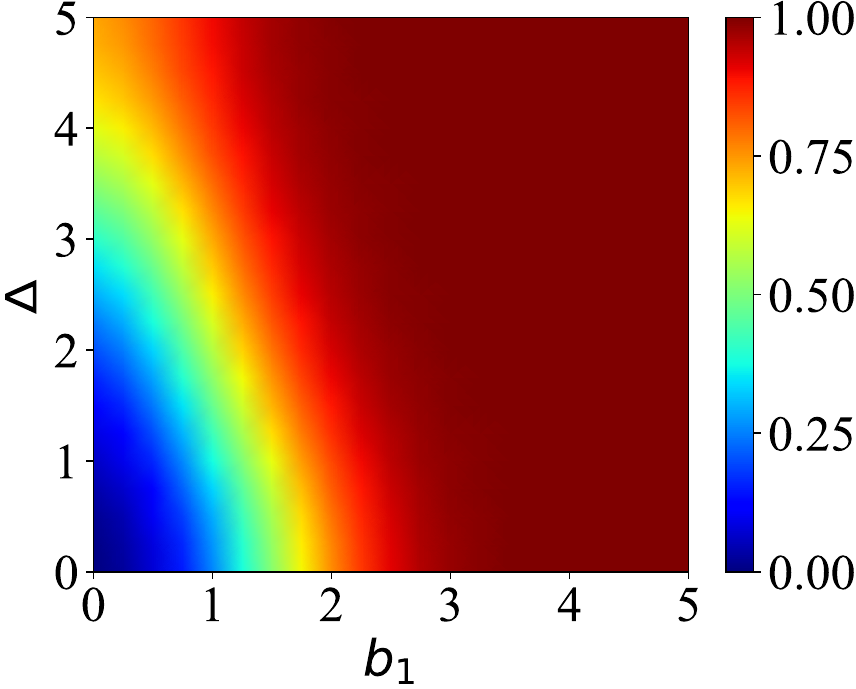}
			\caption{Effect of $b_1$ and $\Delta$}
			\label{fig6b}
		\end{subfigure}
		
		\caption{\textbf{Effect of game parameters on the cooperation frequency.} $\rho_C$ is color coded as shown in the legend. We set the parameters of the Markov decision chain as $\sigma=0.5, \delta=0.1, \gamma=0$ in both panels.  Furthermore, $\Delta=1$ is fixed in panel~(\subref{fig6a}) and $c=1$ is fixed in panel~(\subref{fig6b}).
        }
		\label{fig6}
		
	\end{figure}

    To complete our study, we finally discuss the consequence of the payoff matrix elements on the cooperation level when the parameters of the Markov decision chain are fixed as $\sigma=0.5$, $\delta=0.1$, and $\gamma=0$. Our results are summarized in color-coded heat-maps, shown in Fig.~\ref{fig6}. In agreement with previous studies, an increase in the benefit parameter $b_1$ and a decrease in the cost parameter $c$ result in a higher benefit-to-cost ratio for each game state, which enhances the cooperation frequency. However, unlike previous findings, the parameter $\Delta$, which characterizes the disparity between the game states, also has a significant impact on cooperation level. In particular, the frequency of cooperators grows significantly when $\Delta$ is increased. This result suggests that as the agent’s game state could be harsher or more favorable, agents are more inclined to shift the game state toward the one that is more beneficial for the collective, even if this change involves a certain level of risk.

	\section{Conclusion and Outlook}		\label{Conclu}

    Individual decisions in a social dilemma situation have huge consequences, especially when the environment is active~\cite{barfuss2019deterministic}. In many applications, the environment is adaptive, such as when populations aim to control an epidemic~\cite{abel2021socially,johnson2020slowing,chica2021collective}, manage natural resources~\cite{samuelson1990energy,cumming2018review,van2002central}, or mitigate climate change~\cite{milfont2010global, vesely2020pro, tavoni2011inequality}. Our present theoretical study is an initial step along this research path, where we proposed a Markov decision process based on a bidirectional feedback between game state transitions and individual strategies. In this model, individuals adopt one of the two strategies, cooperation or defection, which drive the transition of game states according to the probability matrices. The extent to which each game state supports cooperative behavior varies, which in turn influences the strategy choices of individuals, forming a dynamic and self-regulating bidirectional feedback between strategies and game states. Based on this assumption, we extended traditional decision rules, particularly those based on the widely used imitation rule. Our model allows individuals to update their strategies across multiple future time steps within a single time step, thus capturing more diverse decision making. In our numerical simulations we used three versions of donation game consisting of different payoff matrices to represent different game states. First, we examined the impact of the Markov decision chain's transition probabilities, the policy size of an individual, and the degree of irrationality on each game state. Due to the bidirectional feedback mechanism in the model, we have analyzed the indirect impact of parameters on the distribution of game states. Specifically, we have explored the effect of the Markov decision chain's transition probabilities, policy size, and degree of irrationality on the convergence of cooperator density within the population. Additionally, we have examined the direct impact of the payoff values and their differences among different game states on the cooperator density. 
    
    Our results indicate that the cooperation level can be increased significantly even if the well-known condition that the benefit–cost ratio must exceed the average degree is not met. The simulation results provide a numerical assessment of how factors such as the rate of environmental change and the degree of environmental heterogeneity influence the spread of cooperation. Notably, this system behavior is more pronounced when the payoff differences and transition rates between various game states are larger. This implies that by adjusting the specific environmental transition rates or by amplifying the degree of environmental differences, the trajectory of cooperation becomes controllable. Our work provides concrete guidance for such regulation through simulation results.
        
    In this study, there are numerous model variations and evolutionary details that warrant further exploration. Furthermore, our work focuses on two-player games, and the conclusions share similarities with past research on both pairwise and multi-player games~\cite{Hilbe2018,he2022persistence}. For instance, the direct influence of cooperation and defection on the environment promotes the emergence of cooperation. Our present model uses donation games with varying payoff parameters. Incorporating other types of social dilemmas game or a mixture of multiple models may yield additional noteworthy conclusions. Furthermore, our model extends the traditional pairwise imitation strategy learning based on the Fermi function, providing a generalized approach. Building on this foundation, further expansion to incorporate diverse evolutionary strategies, as well as combinations of these strategies, could more effectively enhance our understanding about how cooperation emerges and how the diversity of decision-making can be characterized more accurately.

    \section*{CRediT authorship contribution statement}

    \textbf{Chaoyang Luo:} Conceptualization, Methodology, Validation and Writing - Original Draft. \textbf{Yuji Zhang:} Methodology, Validation and Writing - Review \& Editing. \textbf{Minyu Feng:} Conceptualization, Formal analysis, Validation and Writing - Review \& Editing. \textbf{Attila Szolnoki:} Formal analysis, Validation and Writing - Original Draft.

\singlespacing
\printbibliography[title=References]

@article{Acar2008,
  title={Stochastic switching as a survival strategy in fluctuating environments},
  author={Acar, Murat and Mettetal, Jerome T and Van Oudenaarden, Alexander},
  journal={Nat. Genet.},
  volume={40},
  number={4},
  pages={471--475},
  year={2008},
  publisher={Nature Publishing Group US New York}
}

@article{Aming2016,
  title={Evolutionary dynamics of general group interactions in structured populations},
  author={Li, Aming and Broom, Mark and Du, Jinming and Wang, Long},
  journal={Phys. Rev. E},
  volume={93},
  number={2},
  pages={022407},
  year={2016},
  publisher={APS}
}

@article{Feng2023,
  title={An evolutionary game with the game transitions based on the Markov process},
  author={Feng, Minyu and Pi, Bin and Deng, Liang-Jian and Kurths, J{\"u}rgen},
  journal={IEEE Trans. Syst., Man, Cybern. Syst.},
  year={2023},
  publisher={IEEE}
}

@article{Franzenburg2013,
  title={Bacterial colonization of Hydra hatchlings follows a robust temporal pattern},
  author={Franzenburg, S{\"o}ren and Fraune, Sebastian and Altrock, Philipp M and K{\"u}nzel, Sven and Baines, John F and Traulsen, Arne and Bosch, Thomas CG},
  journal={ISME J.},
  volume={7},
  number={4},
  pages={781--790},
  year={2013},
  publisher={Oxford University Press}
}

@article{Hilbe2018,
  title={Evolution of cooperation in stochastic games},
  author={Hilbe, Christian and {\v{S}}imsa, {\v{S}}t{\v{e}}p{\'a}n and Chatterjee, Krishnendu and Nowak, Martin A},
  journal={Nature},
  volume={559},
  number={7713},
  pages={246--249},
  year={2018},
  publisher={Nature Publishing Group UK London}
}

@article{Ohtsuki2006,
  title={A simple rule for the evolution of cooperation on graphs and social networks},
  author={Ohtsuki, Hisashi and Hauert, Christoph and Lieberman, Erez and Nowak, Martin A},
  journal={Nature},
  volume={441},
  number={7092},
  pages={502--505},
  year={2006},
  publisher={Nature Publishing Group UK London}
}

@article{Perc2012,
  title={Evolutionary dynamics of group interactions on structured populations: a review},
  author={Perc, Matja{\v{z}} and G{\'o}mez-Gardenes, Jes{\'u}s and Szolnoki, Attila and Flor{\'\i}a, Luis M and Moreno, Yamir},
  journal={J. R. Soc. Interface},
  volume={10},
  number={80},
  pages={20120997},
  year={2013},
  publisher={The Royal Society}
}

@article{Rankin2007,
  title={The tragedy of the commons in evolutionary biology},
  author={Rankin, Daniel J and Bargum, Katja and Kokko, Hanna},
  journal={Trends Ecol. Evol.},
  volume={22},
  number={12},
  pages={643--651},
  year={2007},
  publisher={Elsevier}
}

@article{Simon2014,
  title={Public goods in relation to competition, cooperation, and spite},
  author={Levin, Simon A},
  journal={Proc. Natl. Acad. Sci. U. S. A.},
  volume={111},
  number={supplement\_3},
  pages={10838--10845},
  year={2014},
  publisher={National Acad Sciences}
}

@article{Su2019,
  title={Evolutionary dynamics with game transitions},
  author={Su, Qi and McAvoy, Alex and Wang, Long and Nowak, Martin A},
  journal={Proc. Natl. Acad. Sci. U. S. A.},
  volume={116},
  number={51},
  pages={25398--25404},
  year={2019},
  publisher={National Acad Sciences}
}

@article{Vasconcelos2017,
  title={Stochastic dynamics through hierarchically embedded Markov chains},
  author={Vasconcelos, V{\'\i}tor V and Santos, Fernando P and Santos, Francisco C and Pacheco, Jorge M},
  journal={Phys. Rev. Lett.},
  volume={118},
  number={5},
  pages={058301},
  year={2017},
  publisher={APS}
}

@article{wang2006,
  title={Memory-based snowdrift game on networks},
  author={Wang, Wen-Xu and Ren, Jie and Chen, Guanrong and Wang, Bing-Hong},
  journal={Phys. Rev. E},
  volume={74},
  number={5},
  pages={056113},
  year={2006},
  publisher={APS}
}

@article{Deonauth2021,
  author={Deonauth, Nakema Y. and Li, Mingchu and Yu, Shuo and Chen, Xiangtai},
  journal={IEEE Trans. Comput. Soc. Syst.},
  title={An Upstream-Reciprocity-Based Strategy for Academic Social Networks Using Public Goods Game},
  year={2021},
  volume={8},
  number={6},
  pages={1417--1426},
  doi={10.1109/TCSS.2021.3085174}
}

@ARTICLE{Xiao2023,
  author={Xiao, Yunpeng and He, Weikang and Yang, Tao and Li, Qian},
  journal={IEEE Trans. Comput. Soc. Syst.},
  title={A Dynamic Information Dissemination Model Based on User Awareness and Evolutionary Games},
  year={2023},
  volume={10},
  number={5},
  pages={2837--2846},
  doi={10.1109/TCSS.2022.3201061}
}

@article{Li2026,
title = {Memory-driven Q-learning model for cooperation in snowdrift game with dynamic behavioral types},
journal = {Appl. Math. Model.},
volume = {149},
pages = {116313},
year = {2026},
issn = {0307-904X},
doi = {10.1016/j.apm.2025.116313},
author = {Xiang Li and Bin Pi and Liang-Jian Deng and Qin Li},
}

@article{zeng2025complex,
  author  = {Zeng, Ziyan and Feng, Minyu and Liu, Pengfei and Kurths, J{\"u}rgen},
  journal = {IEEE Trans. Syst., Man, Cybern. Syst.},
  title   = {Complex Network Modeling With Power-Law Activating Patterns and Its Evolutionary Dynamics},
  year    = {2025},
  volume  = {55},
  number  = {4},
  pages   = {2546--2559},
  doi     = {10.1109/TSMC.2025.3525465}
}

@book{tanimoto2021sociophysics,
  title={Sociophysics approach to epidemics},
  author={Tanimoto, Jun},
  volume={23},
  year={2021},
  publisher={Springer}
}

@ARTICLE{Pi2025Dynamic,
    author={Pi, Bin and Deng, Liang-Jian and Feng, Minyu and Perc, Matjaž and Kurths, Jürgen},
    journal={IEEE Trans. Pattern Anal. Mach. Intell.},
    title={Dynamic Evolution of Complex Networks: A Reinforcement Learning Approach Applying Evolutionary Games to Community Structure},
    year={2025},
    volume={47},
    number={10},
    pages={8563--8582},
    doi={10.1109/TPAMI.2025.3579895}
}

@article{zhang2024limitation,
    title={Limitation of time promotes cooperation in structured collaboration systems},
    author={Zhang, Yichao and Wang, Jiasheng and Wen, Guanghui and Guan, Jihong and Zhou, Shuigeng and Chen, Guanrong and Chatterjee, Krishnendu and Perc, Matja{\v{z}}},
    journal={IEEE Trans. Netw. Sci. Eng.},
    year={2024}
}

@article{zeng2025bursty,
    title={Bursty switching dynamics promotes the collapse of network topologies},
    author={Zeng, Ziyan and Feng, Minyu and Perc, Matja{\v{z}} and Kurths, J{"u}rgen},
    journal={Proc. R. Soc. A},
    volume={481},
    number={2310},
    pages={20240936},
    year={2025}
}

@article{kleshnina2023effect,
    title={The effect of environmental information on evolution of cooperation in stochastic games},
    author={Kleshnina, Maria and Hilbe, Christian and {\v{S}}imsa, {\v{S}}t{\v{e}}p{'a}n and Chatterjee, Krishnendu and Nowak, Martin A},
    journal={Nat. Commun.},
    volume={14},
    number={1},
    pages={4153},
    year={2023}
}

@article{tilman2020evolutionary,
    title={Evolutionary games with environmental feedbacks},
    author={Tilman, Andrew R and Plotkin, Joshua B and Ak{\c{c}}ay, Erol},
    journal={Nat. Commun.},
    volume={11},
    number={1},
    pages={915},
    year={2020}
}

@article{menge2008evolutionary,
    title={Evolutionary tradeoffs can select against nitrogen fixation and thereby maintain nitrogen limitation},
    author={Menge, Duncan NL and Levin, Simon A and Hedin, Lars O},
    journal={Proc. Natl. Acad. Sci. U.S.A.},
    volume={105},
    number={5},
    pages={1573--1578},
    year={2008}
}

@article{grman2012ecological,
    title={Ecological specialization and trade affect the outcome of negotiations in mutualism},
    author={Grman, Emily and Robinson, Todd MP and Klausmeier, Christopher A},
    journal={Am. Nat.},
    volume={179},
    number={5},
    pages={567--581},
    year={2012}
}

@article{su2025evolutionary,
    title={Evolutionary dynamics of behavioral motivations for cooperation},
    author={Su, Qi and Stewart, Alexander J},
    journal={Nat. Commun.},
    volume={16},
    number={1},
    pages={4023},
    year={2025}
}

@article{tarnita2025reconciling,
title={Reconciling ecology and evolutionary game theory or “When not to think cooperation”},
author={Tarnita, Corina E and Traulsen, Arne},
journal={Proc. Natl. Acad. Sci. U.S.A.},
volume={122},
number={14},
pages={e2413847122},
year={2025}
}

@article{barfuss2019deterministic,
  title={Deterministic limit of temporal difference reinforcement learning for stochastic games},
  author={Barfuss, Wolfram and Donges, Jonathan F and Kurths, J{\"u}rgen},
  journal={Phys. Rev. E},
  volume={99},
  number={4},
  pages={043305},
  year={2019}
}

@article{he2022persistence,
  title={The persistence and transition of multiple public goods games resolves the social dilemma},
  author={He, Jialu and Wang, Jianwei and Yu, Fengyuan and Chen, Wei and Xu, Wenshu},
  journal={Appl. Math. Comput.},
  volume={418},
  pages={126858},
  year={2022}
}

@article{abel2021socially,
  title={Socially optimal mistakes? Debiasing COVID-19 mortality risk perceptions and prosocial behavior},
  author={Abel, Martin and Byker, Tanya and Carpenter, Jeffrey},
  journal={J. Econ. Behav. Organ.},
  volume={183},
  pages={456--480},
  year={2021}
}

@article{johnson2020slowing,
  title={Slowing COVID-19 transmission as a social dilemma: Lessons for government officials from interdisciplinary research on cooperation},
  author={Johnson, Tim and Dawes, Christopher and Fowler, James and Smirnov, Oleg},
  journal={J. Behav. Public Adm.},
  volume={3},
  number={1},
  year={2020}
}

@article{chica2021collective,
  title={A collective risk dilemma for tourism restrictions under the COVID-19 context},
  author={Chica, Manuel and Hern{\'a}ndez, Juan M and Bulchand-Gidumal, Jacques},
  journal={Sci. Rep.},
  volume={11},
  number={1},
  pages={5043},
  year={2021}
}

@article{samuelson1990energy,
  title={Energy conservation: A social dilemma approach},
  author={Samuelson, Charles D},
  journal={Soc. Behav.},
  volume={5},
  number={4},
  pages={207--230},
  year={1990}
}

@article{cumming2018review,
  title={A review of social dilemmas and social-ecological traps in conservation and natural resource management},
  author={Cumming, Graeme S},
  journal={Conserv. Lett.},
  volume={11},
  number={1},
  pages={e12376},
  year={2018}
}

@article{van2002central,
  title={Central, individual, or collective control? Social dilemma strategies for natural resource management},
  author={Van Vugt, Mark},
  journal={Am. Behav. Sci.},
  volume={45},
  number={5},
  pages={783--800},
  year={2002}
}

@article{tavoni2011inequality,
  title={Inequality, communication, and the avoidance of disastrous climate change in a public goods game},
  author={Tavoni, Alessandro and Dannenberg, Astrid and Kallis, Giorgos and L{\"o}schel, Andreas},
  journal={Proc. Natl. Acad. Sci. U.S.A.},
  volume={108},
  number={29},
  pages={11825--11829},
  year={2011}
}

@article{milfont2010global,
  title={Global warming, climate change and human psychology},
  author={Milfont, Taciano L},
  journal={Psychol. Approaches Sustain.: Curr. Trends Theory Res. Pract.},
  volume={19},
  pages={42},
  year={2010}
}

@article{vesely2020pro,
  title={Pro-environmental behavior as a signal of cooperativeness: Evidence from a social dilemma experiment},
  author={Vesely, Stepan and Kl{\"o}ckner, Christian A and Brick, Cameron},
  journal={J. Environ. Psychol.},
  volume={67},
  pages={101362},
  year={2020}
}

\end{document}